\documentclass[twocolumn,notitlepage,prl,superscriptaddress,longbibliography]{revtex4-2}
\usepackage{amsfonts}
\usepackage{textcomp}
\usepackage{times}
\usepackage{graphicx}
\usepackage{float}
\usepackage{latexsym,amsmath,amssymb,bm,euscript}
\usepackage{color}
\usepackage{subfigure}
\usepackage{epstopdf}
\usepackage[colorlinks=true,linkcolor=blue,citecolor=blue,urlcolor=blue]{hyperref}
\usepackage{hyperref}
\usepackage{soul}
\usepackage[normalem]{ulem}
\usepackage{mathrsfs}
\usepackage{amsmath}
\usepackage{xspace}
\usepackage{natbib}
\usepackage{ulem}
\usepackage{etoolbox}
\usepackage{tikz}

\usepackage{xcolor}
\definecolor{orange}{rgb}{1,0.5,0}

\usepackage[normalem]{ulem}
\usepackage{comment}

\begin{document}

\title{Emergent Superconductivity and Competing Charge Orders in Hole-Doped Square-Lattice $t$-$J$ Model}

\author{Xin Lu}\thanks{Both authors contributed equally to this work.}
\affiliation{School of Physics, Beihang University, Beijing 100191, China}

\author{Feng Chen}\thanks{Both authors contributed equally to this work.}
\affiliation{Department of Physics and Astronomy, California State University Northridge, Northridge, California 91330, USA}

\author{W. Zhu}
\affiliation{School of Science, Westlake University, Hangzhou 310024, China, and \\
Institute of Natural Sciences, Westlake Institute of Advanced Study, Hangzhou 310024, China, and \\
Key Laboratory for Quantum Materials of Zhejiang Province, Westlake University, Hangzhou 310024, China}

\author{D. N. Sheng}
\email{donna.sheng1@csun.edu}
\affiliation{Department of Physics and Astronomy, California State University Northridge, California 91330, USA}

\author{Shou-Shu Gong}
\email{shoushu.gong@gbu.edu.cn}
\affiliation{School of Physical Sciences, Great Bay University, Dongguan 523000, China, and \\
Great Bay Institute for Advanced Study, Dongguan 523000, China}

\date{\today}

\begin{abstract}
The square-lattice Hubbard and closely related $t$-$J$ models are considered as basic paradigms for understanding strong correlation effects and unconventional superconductivity (SC).
Recent large-scale density matrix renormalization group simulations on the extended $t$-$J$ model have identified $d$-wave SC on the electron-doped side (with the next-nearest-neighbor hopping $t_2>0$) but a dominant charge density wave (CDW) order on the hole-doped side ($t_2<0$), which is inconsistent with the SC of hole-doped cuprate compounds. 
We re-examine the ground-state phase diagram of the extended $t$-$J$ model by employing the state-of-the-art density matrix renormalization group calculations with much enhanced bond dimensions, allowing more accurate determination of the ground state.
On six-leg cylinders, while different CDW phases are identified on the hole-doped side for the doping range $\delta= 1/16-1/8$, a SC phase emerges at a lower doping regime, with algebraically decaying pairing correlations and $d$-wave symmetry.
On the wider eight-leg systems, the $d$-wave SC also emerges on the hole-doped side at the optimal $1/8$ doping, 
demonstrating the winning of SC over CDW by increasing the system width.  
Our results not only suggest a new path to SC in general $t$-$J$ models through weakening the competing charge orders, but also provide a unified understanding on the SC of both hole- and electron-doped cuprate superconductors.
\end{abstract}

\maketitle

{\it Introduction.---}
Understanding the mechanism of unconventional superconductivity (SC) in cuprates is a major challenge of condensed matter physics~\cite{Keimer_nature_2015,Proust_ARCMP_2019}.
Soon after the discovery of cuprate superconductors, the resonating valence bond (RVB) theory~\cite{Anderson_Science_1987} was proposed to describe unconventional SC. 
The square Hubbard (with a large $U$) and closely related $t$-$J$  models are considered as the minimum models~\cite{Keimer_nature_2015,Proust_ARCMP_2019,Anderson_Science_1987,Arovas_ARCMP_2022,Sachdev_RMP_2003,Anderson_2004,Lee_RMP_2006,Masao_RPP_2008} to realize unconventional SC, which have attracted intense explorations~\cite{Anderson_PRL_1987,Anderson_2004,Lee_RMP_2006,Kivelson_PRB_1987,Anderson_2004,Masao_RPP_2008,Weng_PRB_1997,senthil_2000,Kaul_NP_2007,weng_2011}.
However, it remains illusive if these models can describe the SC of cuprates.
In the presence of strong correlations, analytical solutions are not controlled, while numerical studies in the relevant regime~\cite{White_PRL_1998,White_PRB_1999,White_PRL_2003,Hager_PRB_2005,Corboz_PRL_2014,Simons_PRX_2015,Ehlers_PRB_2017,Garnet_Science_2017,EWHuang_Science_2017,Ido_PRB_2018,Jiang_PRB_2018,Corboz_PRB_2019,QinMingPu_PRX_2020,Xu_PRR_2022,Martins_PRB_2001,Sorella_PRL_2002,Raczkowski2007,Capello2008,Himeda_PRL_2002,Shih_PRL_2004,White_PRB_2009,Eberlein_PRB_2014,Dodaro_PRB_2017,Jiang_Science_2019,Chung_PRB_2020,Jiang_PRR_2020,White_PNAS_2021,Jiang_PRL_2021,Gong_PRL_2021,White_PRB_2022,Wietek_PRL_2022,Wietek_PRX_2021,Qu2022,Jiang2023} are also extremely difficult in determining the ground state due to the extensive entanglement and low-energy excitations associated with competing spin and charge degrees of freedom. 
In recent years, numerical simulations have reached a possible consensus on the ground states of the pure large-$U$ Hubbard and $t$-$J$ models near the optimal doping, which is the stripe phase~\cite{White_PRL_1998,White_PRB_1999,White_PRL_2003,Hager_PRB_2005,Corboz_PRL_2014,Simons_PRX_2015,Ehlers_PRB_2017,Garnet_Science_2017,EWHuang_Science_2017,Ido_PRB_2018,Jiang_PRB_2018,Corboz_PRB_2019,QinMingPu_PRX_2020,Xu_PRR_2022} characterized by a charge density wave (CDW) order coexisting with $\pi$-phase shifted antiferromagnetic domains, accompanied by exponentially decaying SC correlation.

On the other hand, the Fermi surface topology identified experimentally for cuprates indicates the importance of a small next-nearest-neighbor hopping $t_2$~\cite{Damascelli_RMP_2003}, with the sign of $t_2$ modeling the hole- ($t_2<0$) and electron-doped ($t_2>0$) cuprates, respectively~\cite{belinicher_generalized_1996}. 
Numerical studies on four-leg Hubbard and $t$-$J$ models find that introducing either positive or negative $t_2$ can lead to the coexistence of quasi-long-range SC and CDW orders~\cite{Dodaro_PRB_2017,Jiang_Science_2019,Chung_PRB_2020,Jiang_PRR_2020}.
To improve our understanding of how these orders evolve toward two dimensions (2D), recent density matrix renormalization group (DMRG) studies on six- and eight-leg $t$-$J$ model (with the nearest-neighbor hopping $t_1>0$) have identified a robust $d$-wave SC with suppressed CDW at $t_2 > 0$~\cite{Jiang_PRL_2021,Gong_PRL_2021,White_PNAS_2021}, giving insights into the SC of electron-doped cuprates.
For $t_2 < 0$, the stripe order appears to win over SC near the optimal doping~\cite{White_PNAS_2021,White_PRB_2022,lu2023sign}, in sharp contrast with hole-doped cuprates~\cite{Scalapino_RMP_2012}. 
However, while accurate DMRG simulations have been applied to six-leg ladders~\cite{Gong_PRL_2021, Jiang_PRL_2021,lu2023sign}, large-bond-dimension simulations are absent for eight-leg systems, which leaves the true nature of the ground state of the hole-doped $t$-$J$ model an open question.

\begin{figure}
   \includegraphics[width=0.485\textwidth,angle=0]{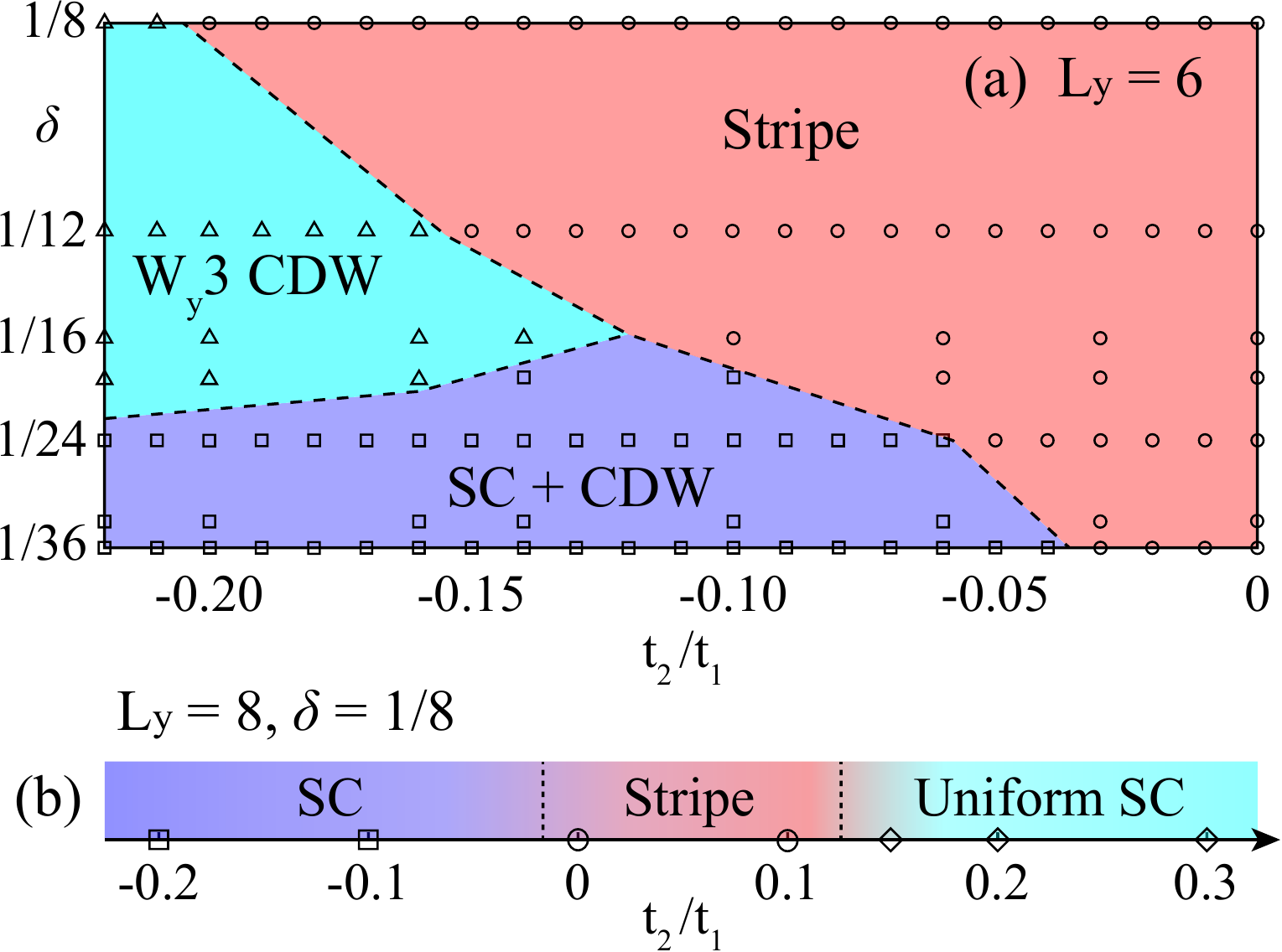}
   \caption{\label{Geo-Pha2}
   Quantum phase diagrams of the $t_1$-$t_2$-$J_1$-$J_2$ model at different system widths.
   (a) $L_y=6$ cylinder with $-0.22 \le t_2 /t_1 \le 0$ and $1/36\le \delta \le1/8$. We identify a stripe phase, a $\mathrm{W_{y}3}$ CDW phase, and a SC + CDW phase with coexisted $d$-wave SC and a weak CDW. (b) $L_y=8$ cylinder with $-0.2 \le t_2 /t_1 \le 0.3$ at $\delta =1/8$. We identify two SC phases and a stripe phase. The hole-doped SC phase at $t_2/t_1 < 0$ has a weak or vanishing CDW order. Pairing correlations in the $L_y = 8$ stripe phase show a slow increase with bond dimension, but its tendency to develop a quasi-long-range SC order cannot be pinned down within our currently accessible bond dimensions. The symbols denote the parameter points that we have calculated. The same SC phases on both six- and eight-leg systems are obtained in our model with $(t_2/t_1)^2 = J_2/J_1$ and the $t_1$-$t_2$-$J_1$ model with $t_1/J_1=2.5$ and $3.0$ (see Fig.~\ref{Com_SC2} and SM~\cite{SM}).}
\end{figure}

In this Letter, we study the phase diagram of the hole-doped $t$-$J$ model and examine the interplay between SC and CDW through accurate DMRG calculations. 
By tuning the doping level $\delta$ and hopping ratio $t_{2}/t_{1}$ on six-leg system, we identify the dominance of CDW phases at $\delta=1/16-1/8$.
However, the SC and weak CDW can coexist at lower doping region $\delta=1/24-1/36$ [Fig.~\ref{Geo-Pha2}(a)], where pairing correlations show the $d$-wave symmetry and slow power-law decay with the exponent $K_{\mathrm{sc}} \lesssim 1$.
Importantly, we observe dominant quasi-long-range SC order at the optimal doping ($\delta=1/8$) on eight-leg cylinder [Fig.~\ref{Geo-Pha2}(b)]. 
On the electron-doped side ($t_2>0$), we confirm the existence of a robust uniform $d$-wave SC in agreement with previous studies~\cite{White_PNAS_2021,White_PRB_2022}. On the hole-doped side ($t_2<0$), we observe the remarkable emergence of SC with weak or vanishing CDW order in our large-bond-dimension simulation, with power-law decaying pairing correlations ($K_{\rm sc} < 2$).
Furthermore, we confirm the robustness of these SC phases at different model parameters.
Our work suggests that the $t$-$J$ model may offer a unified framework for understanding the unconventional SC for both electron- and hole-doped cuprates.

{\it Model and method.---}
The Hamiltonian of the extended $t$-$J$ model is defined as
\begin{equation}\label{H-tJ}
  H = -\sum_{\{ij\},\sigma}t_{ij}(\hat{c}^{\dagger}_{i,\sigma} \hat{c}_{j,\sigma} + {\rm H.c.})  + \sum_{\{ij\}} J_{ij} (\hat{\bf S}_i \cdot \hat{\bf S}_j - \frac{1}{4} \hat{n}_i \hat{n}_j),  \nonumber
\end{equation}
where ${\hat{c} }_{i\sigma }^{\dagger }$ (${\hat{c} }_{i\sigma }$) is the creation (annihilation) operator of the electron with spin $\sigma$ ($\sigma = \pm 1/2$) on site $i=(x_i,y_i)$, ${\hat{\bf S} }_i$ is the spin-$1/2$ operator, and ${\hat{n} }_i =\sum_{\sigma } {\hat{c} }_{i\sigma }^{\dagger } {\hat{c} }_{i\sigma }$ is the electron number operator.
The Hilbert space for each site is constrained by no double occupancy. 
We consider the nearest-neighbor and next-nearest-neighbor hoppings ($t_1$ and $t_2$) and spin interactions ($J_1$ and $J_2$).
We choose $J_1=1.0$ and set $t_1/J_1= 3.0$ to make a connection to the corresponding Hubbard model with $U/t=12$~\cite{Misumi_PRB_2017}.
The length and width of the lattice are denoted as $L_x$ and $L_y$, giving total site number $N = L_x \times L_y$.
The doping ratio $\delta$ is defined as $\delta = N_h / N$ ($N_h$ is the number of doped holes).
We focus on the doping regime $1/36\le \delta \le 1/8$ on six-leg cylinders and $\delta=1/8$ on eight-leg cylinders, and tune $t_2/t_1$ with fixed relation ${\left(t_2 /t_1 \right)}^2=J_2 /J_1$~\cite{Jiang_PRL_2021, Gong_PRL_2021}.
We also examine the SC phases in the $t_1$-$t_2$-$J_1$ model with $t_1/J_1 = 2.5, 3.0$, as shown in Fig.~\ref{Com_SC2}.

We solve the ground state of the system by DMRG~\cite{White_PRL_1992} calculations with ${\rm SU(2)} \otimes {\rm U(1)}$ symmetry implemented~\cite{I.P.McCulloch_2002}.
We study cylindrical systems with open and periodic boundary conditions along the axial ($x$) and circumferential ($y$) directions respectively, and keep the bond dimensions of $\rm SU(2)$ multiplets up to $D = 15 000$ for six-leg and $28 000$ for eight-leg systems, equivalent to about 45 000 and 84 000 U(1) states, respectively, which ensure accurate results with the truncation error less than $1.2 \times 10^{-6}$ for six-leg and $2.5\times {10}^{-5}$ for eight-leg systems [see Supplemental Materials (SM) for more details~\cite{SM}].

{\it Quantum phase diagram.---} 
Our results are summarized in the phase diagram Fig.~\ref{Geo-Pha2} as a function of hopping ratio $t_2 /t_1$ and doping level $\delta$. 
For six-leg system with $-0.22 \le t_2 / t_1 \le 0$ [Fig.~\ref{Geo-Pha2}(a)], we identify two charge ordered phases: a stripe phase with wave vector $Q=(3\pi\delta,0)$ and a $\mathrm{W_{y}3}$ CDW phase with $Q=(6\pi\delta,2\pi/3)$ (see SM for the results of the $\mathrm{W_{y}3}$ state~\cite{SM}), which shares a similar charge density distribution with the W3 phase found in the $t_1$-$t_2$-$J_1$ model~\cite{White_PNAS_2021}.
Strikingly, below $\delta=1/18$, we find a quasi-long-range SC order ($K_{\mathrm{sc}} \lesssim 1$) coexisting with a weak CDW. 

For the eight-leg system with $-0.2 \le t_2 / t_1 \le 0.3$ at $\delta = 1/8$ [Fig.~\ref{Geo-Pha2}(b)], a robust $d$-wave SC order emerges for $t_2 / t_1 \gtrsim 0.12$ with a uniform charge density distribution, which is similar to the uniform SC phase found on six-leg cylinder~\cite{Gong_PRL_2021}.
This uniform SC phase may extend to larger $t_2/t_1$ regime~\cite{Jiang_PRL_2021,Jiang_Kivelson_Lee_2023} and persist in 2D limit. 
Remarkably, the quasi-long-range SC order is also observed on the hole-doped side for $t_2/t_1 \lesssim -0.05$, which exhibits a very weak or vanishing charge order. The SC power exponent $K_{\mathrm{sc}}<2$ indicates a divergent SC susceptibility at zero-temperature limit. 
This result contradicts a previous work studying a similar $t_1$-$t_2$-$J_1$ model that claims the absence of SC at $t_2 < 0$~\cite{White_PNAS_2021}, which may be attributed to the existence of competing charge ordered states in low-energy regime.
In our calculation, extremely large bond dimensions are used for reaching convergence and identifying the emergence of SC.
For both six- and eight-leg systems at hole doping, SC emerges through suppressing charge order.


\begin{figure}
   \includegraphics[width=0.483\textwidth,angle=0]{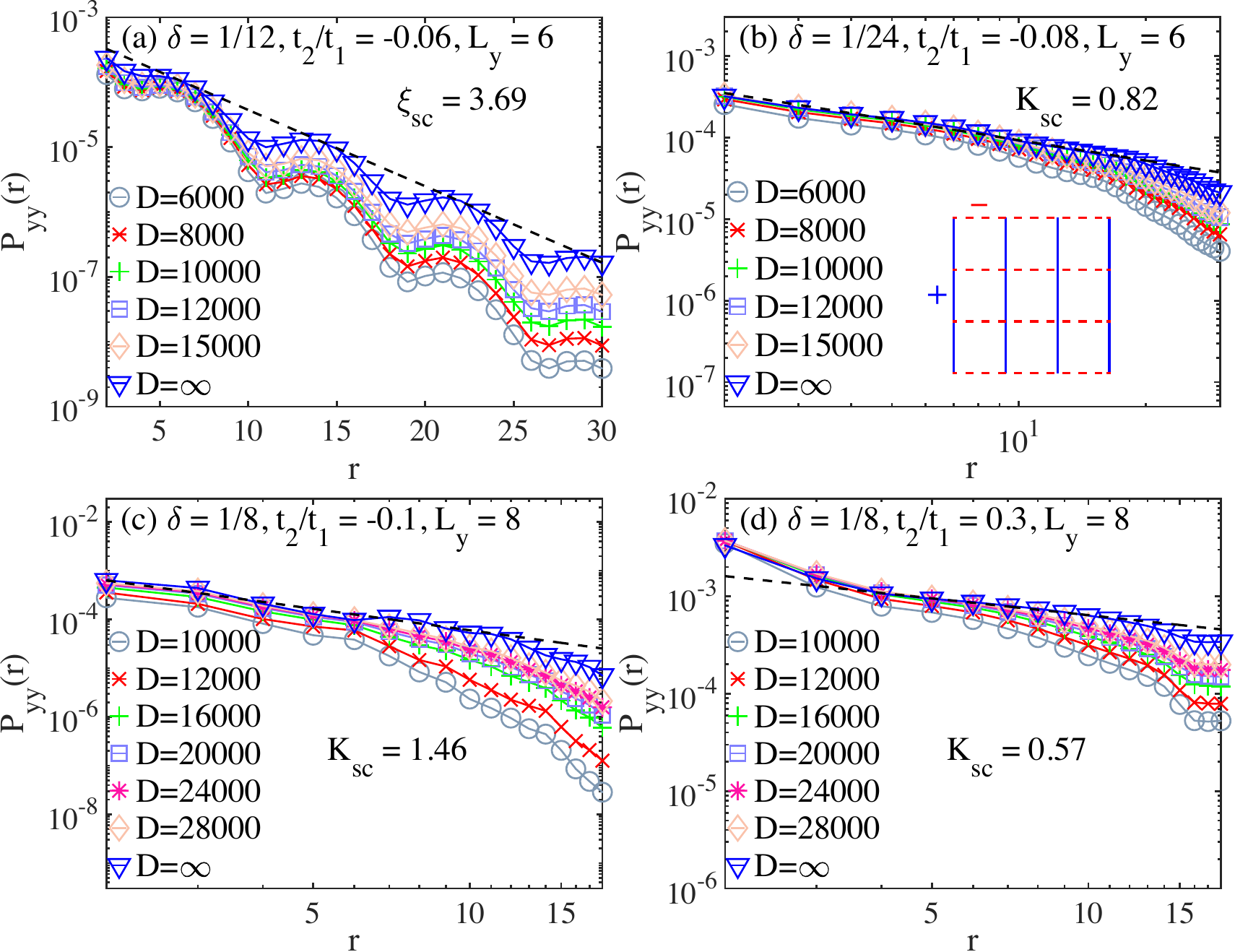}
   \caption{\label{SC_pairing}
   SC pairing correlation functions.
   (a) Semilogarithmic plot of the pairing correlations $P_{yy}(r)$ at different bond dimensions in the stripe phase at $L_y = 6$. The correlation length $\xi_{\mathrm{sc}}$ is obtained by exponential fitting. (b) doublelogarithmic plot of $P_{yy}(r)$ in the SC + CDW phase on six-leg cylinder. The dash line represents the algebraic fitting of the data extrapolated to $D\to \infty$. The power exponent $K_{\mathrm{sc}}\simeq 0.82$ characterizes a quasi-long-range SC order. The inset shows the $d$-wave pairing symmetry. (c) and (d) are similar plots in the hole-doped SC ($t_2 < 0$) and electron-doped uniform $d$-wave SC phases ($t_2 > 0$) on eight-leg cylinders, both with $K_{\mathrm{sc}}<2$ indicating the divergence of SC susceptibilities~\cite{Arrigoni_PRB_2004}.
   }
\end{figure}

{\it SC pairing correlation.---}
We examine SC by the dominant spin-singlet pairing correlations $P_{\alpha, \beta }(\mathbf{r}) = \langle {\hat{\Delta}}_{\alpha }^{\dagger } ({\mathbf{r}}_0) {\hat{\Delta}}_{\beta }({\mathbf{r}}_0 +\mathbf{r}) \rangle$, where the pairing operator is defined as ${\hat{\Delta} }_{\alpha } \left(\mathbf{r}\right)=\left({\hat{c} }_{\mathbf{r}\uparrow } {\hat{c} }_{\mathbf{r}+{\mathbf{e}}_{\alpha } \downarrow } -{\hat{c} }_{\mathbf{r}\downarrow } {\hat{c} }_{\mathbf{r}+{\mathbf{e}}_{\alpha } \uparrow } \right)/\sqrt{2}$ and ${\mathit{\mathbf{e}}}_{\alpha = x,y}$ denote the unit vectors along the $x$ and $y$ directions. 
Since the wave function in DMRG calculation is represented as a matrix product state, correlation functions usually decay exponentially at finite bond dimensions~\cite{RevModPhys.77.259}. 
We  make the bond dimension scaling to demonstrate the true nature of correlations at $D\to \infty$ (see Fig.~\ref{SC_pairing} and SM~\cite{SM}).

We first examine pairing correlations on six-leg systems.
In the stripe phase represented by $t_2/t_1=-0.06$ and $\delta=1/12$  [Fig.~\ref{SC_pairing}(a)], the pairing correlation $P_{yy}(r)$ follows an exponential decay $P_{yy}(r)\sim \mathrm{exp}\left(-r/\xi_{\mathrm{sc}} \right)$ with $\xi_{\mathrm{sc}} \simeq 3.69$ after the extrapolation to $D\to \infty$. 
In the SC + CDW phase, as shown in Fig.~\ref{SC_pairing}(b) for $t_2/t_1=-0.08$, $\delta=1/24$, $P_{yy}(r)$ increases drastically compared with that in the stripe phase and exhibits an algebraic decay $P_{yy}(r) \sim r^{-K_{\mathrm{sc}} }$ with $K_{\mathrm{sc}} \simeq 0.82$, characterizing a quasi-long-range SC order. 
We also confirm that other pairing correlations satisfy $P_{yy}(r) \simeq -P_{yx}(r) \simeq P_{xx} (r)$, in accordance with the $d$-wave symmetry illustrated in the inset of Fig.~\ref{SC_pairing}(b) rather than the plaquette $d$-wave symmetry found in the four-leg Hubbard model at $t_2 < 0$~\cite{Chung_PRB_2020}.

To further investigate whether SC can emerge on wider systems, we extensively simulate the eight-leg cylinder at $\delta=1/8$, which is more relevant to the experiments of cuprates. 
For $t_2/t_1=-0.1$ [Fig.~\ref{SC_pairing}(c)], the pairing correlations at long distance grow rapidly with bond dimension. 
The extrapolated results at $D \rightarrow \infty$ can be fitted by a power-law decay with $K_{\mathrm{sc}} \simeq 1.46$, demonstrating an emergent quasi-long-range SC order. 
In the uniform SC phase at $t_{2}>0$ [Fig.~\ref{SC_pairing}(d)], pairing correlation exhibits a slow algebraic decay with a small exponent $K_{\rm sc} \simeq 0.57$ characterizing a robust SC phase. 
We have also checked the triplet pairing correlations in both SC phases on eight-leg systems.
While the $p$-wave symmetry can appear at $t_2 > 0$, the  corresponding pairing correlations always decay very fast, indicating the absence of triplet SC order~\cite{SM}.


\begin{figure}
   \includegraphics[width=0.483\textwidth,angle=0]{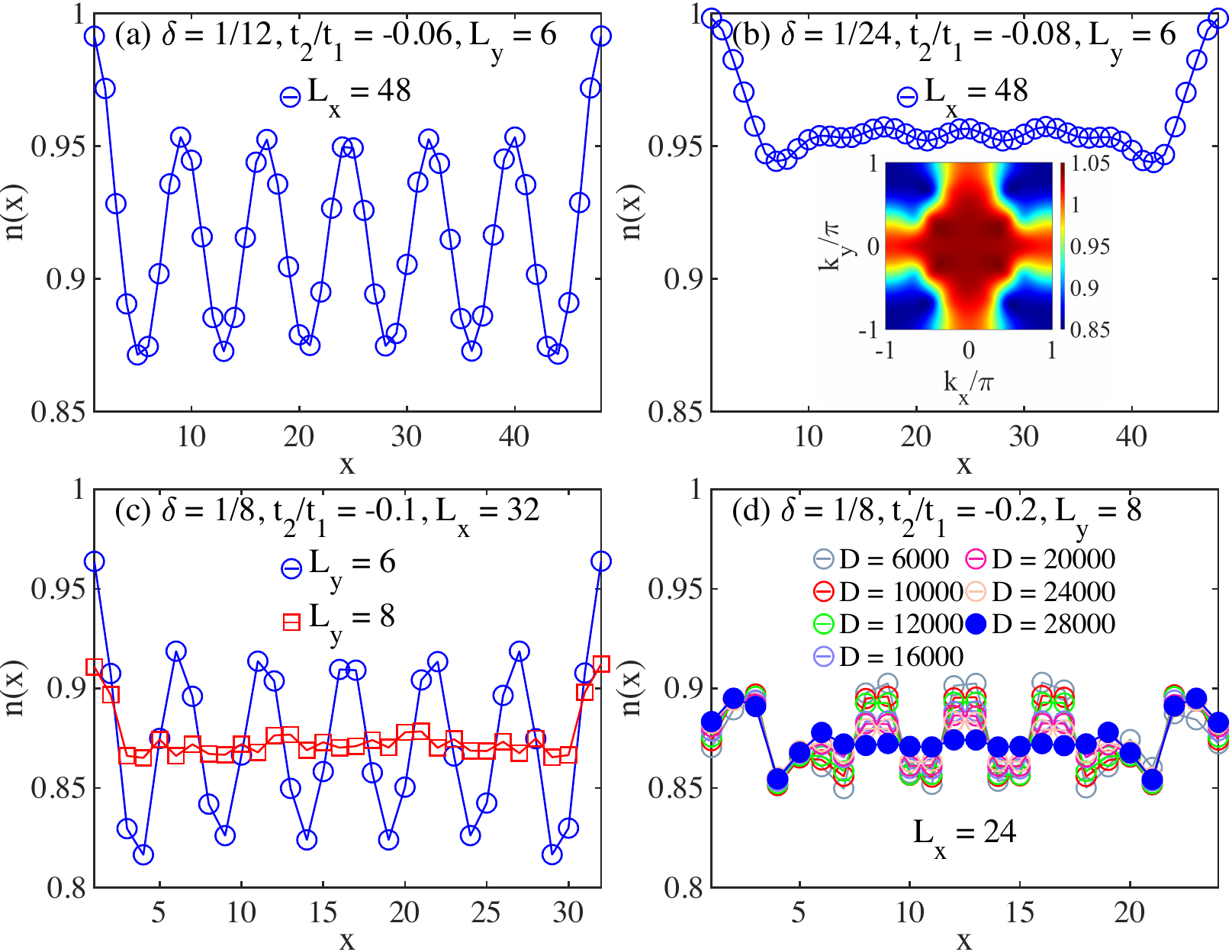}
   \caption{\label{nk_CDW}
   Charge density profiles $n(x)$ in the (a) stripe phase and (b) SC + CDW phase on six-leg cylinders with $L_x=48$. The inset of (b) shows the corresponding electron momentum distribution $n(\bf k)$. (c) Comparing $n(x)$ in the SC phase on $L_y = 8$ and stripe phase on $L_y = 6$ at $t_2/t_1=-0.1$ and $\delta = 1/8$, obtained with $D=24 000$ and $15 000$, respectively. (d) $n(x)$ in the SC phase of eight-leg cylinder at $t_2/t_1=-0.2$, $\delta = 1/8$ obtained by different bond dimensions.
   }
\end{figure}

{\it Charge density distribution.---} 
Except in the $\mathrm{W_{y}3}$ phase, the converged charge density distributions are uniform along the $y$ direction, and we show the averaged charge density for each column as $n(x) = \sum_{y=1}^{L_y} \langle {\hat{n} }_{x,y} \rangle / L_y$ in Fig.~\ref{nk_CDW}. 
For six-leg systems, we find the CDW wavelength $\lambda \simeq 4/(L_{y}\delta)$ in the stripe phase [Fig.~\ref{nk_CDW}(a)], corresponding to four holes on average for each CDW unit.
In the SC + CDW phase, $\lambda \simeq 2/(L_{y}\delta)$ indicates two holes per CDW unit [Fig.~\ref{nk_CDW}(b)]. 
Significantly, the oscillation amplitude of $n(x)$ (i.e. charge order) is much weaker than that in the stripe phase shown in Fig.~\ref{nk_CDW}(a). 
The momentum distribution $n(\mathbf{k}) = (1/N) \sum_{i,j,\sigma} \langle {\hat{c} }_{i,\sigma}^{\dagger } {\hat{c} }_{j,\sigma} \rangle e^{i\mathbf{k}\cdot \left({\mathbf{r}}_i -{\mathbf{r}}_j \right)}$ in the SC + CDW phase [the inset of Fig.~\ref{nk_CDW}(b)] exhibits the unenclosed Fermi surface topology around ${\bf k} = (\pm \pi,0)$ and $(0,\pm \pi)$ in agreement with that observed in the ARPES measurement of hole-doped cuprates~\cite{Damascelli_RMP_2003,Hossain_nphys_2008,plate_fermi_2005}, which is distinctly different from the topology for electron doping at $t_2 > 0$~\cite{Gong_PRL_2021,White_PNAS_2021}, where the Fermi surface forms a closed pocket around ${\bf k} = (0,0)$.

A natural question is how the charge order evolves toward 2D limit.
Crucially, we find that the strong CDW in the stripe phase for $L_y=6$ can be significantly suppressed on wider system, as shown in Fig.~\ref{nk_CDW}(c).
The quite weak charge density oscillation for $L_y = 8$ is similar to that of the SC + CDW phase on six-leg cylinders [Fig.~\ref{nk_CDW}(b)], which is accompanied with the emergent quasi-long-range SC order [Fig.~\ref{SC_pairing}(c)].
In Fig.~\ref{nk_CDW}(d) for $t_2/t_1=-0.2$, one can find the charge distribution is gradually transformed from a CDW-like pattern to a nearly uniform one with growing bond dimension, demonstrating an extremely weak or vanishing charge order in the hole-doped SC phase and the importance of a large bond dimension for reaching the true ground state (see SM~\cite{SM}).

\begin{figure}
   \includegraphics[width=0.483\textwidth,angle=0]{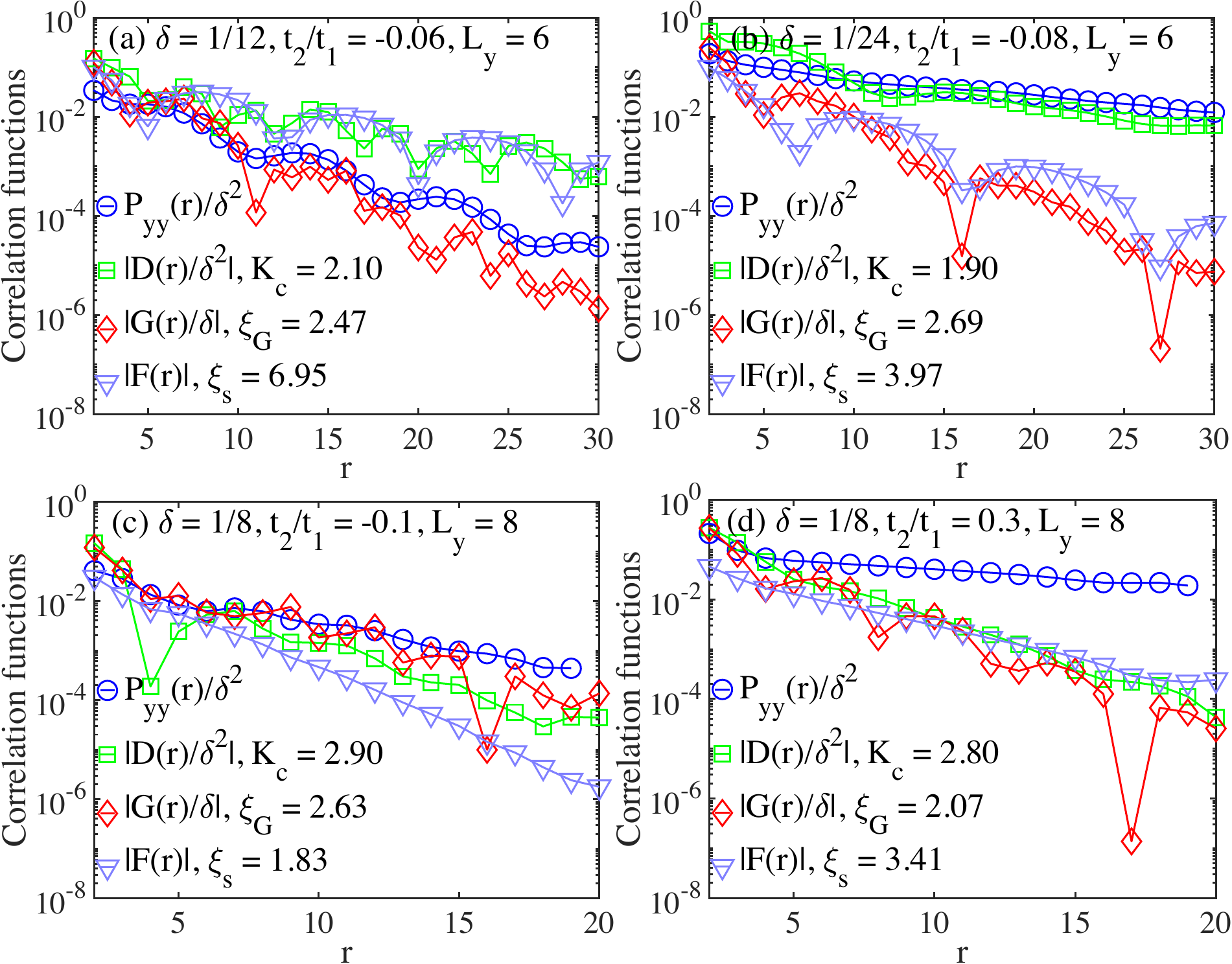}
   \caption{\label{Correlations_Stripe_SC}
   Correlations in different phases. The data are those extrapolated to $D\rightarrow \infty$. Comparison of pairing correlation $P_{yy}(r)$, charge density correlation $D(r)$, single-particle Green's function $G(r)$, and spin correlation $F(r)$ for (a) stripe phase at $L_y = 6$, (b) SC + CDW phase at $L_y = 6$, (c) hole-doped SC phase at $L_y = 8, \delta = 1/8$, and (d) uniform $d$-wave SC phase at $L_y = 8, \delta = 1/8$.
   The correlations are rescaled by $\delta$ to make a direct comparison. The power exponent $K$ and correlation length $\xi$ are obtained by algebraic and exponential fittings, respectively (see the details in SM~\cite{SM}).}
\end{figure}

{\it Correlation functions.---}
In Fig.~\ref{Correlations_Stripe_SC}, we further compare correlation functions in each phase.
While all the correlations are presented in the semilogarithmic scale, the exponents $K$ and correlation lengths $\xi$ are obtained by power-law and exponential fittings, respectively~\cite{SM}.
For the stripe phase on six-leg cylinders [Fig.~\ref{Correlations_Stripe_SC}(a)], while the single-particle Green’s function $G(r) = \langle \sum_{\sigma} \hat{c}_{x,y,\sigma }^{\dagger } \hat{c}_{x+r,y,\sigma } \rangle$ and pairing correlation appear to decay exponentially~\cite{SM}, the intertwined density correlation $D(r) = \langle {\hat{n} }_{x,y} {\hat{n} }_{x+r,y} \rangle - \langle {\hat{n} }_{x,y} \rangle \langle {\hat{n} }_{x+r,y} \rangle$ and spin correlation $F(r)=\langle {\hat{\mathbf{S}}}_{x,y} \cdot {\hat{\mathbf{S}}}_{x+r,y} \rangle$ are more dominant at long distance.
In contrast, in the SC + CDW [Fig.~\ref{Correlations_Stripe_SC}(b)], hole-doped SC [Fig.~\ref{Correlations_Stripe_SC}(c)], and uniform $d$-wave SC phases [Fig.~\ref{Correlations_Stripe_SC}(d)], pairing correlations are dominant over other correlations at long distance.
Furthermore, on eight-leg systems, $G(r)$ and $F(r)$ show exponential decay with short correlation lengths at $t_2/t_1=0.3$, which is consistent with the DMRG results of the same model at $t_2/t_1\approx 0.5$ corresponding to doping either the $J_1$-$J_2$ spin liquid or valence bond solid~\cite{Jiang_Kivelson_Lee_2023}.

\begin{figure}
   \includegraphics[width=0.485\textwidth,angle=0]{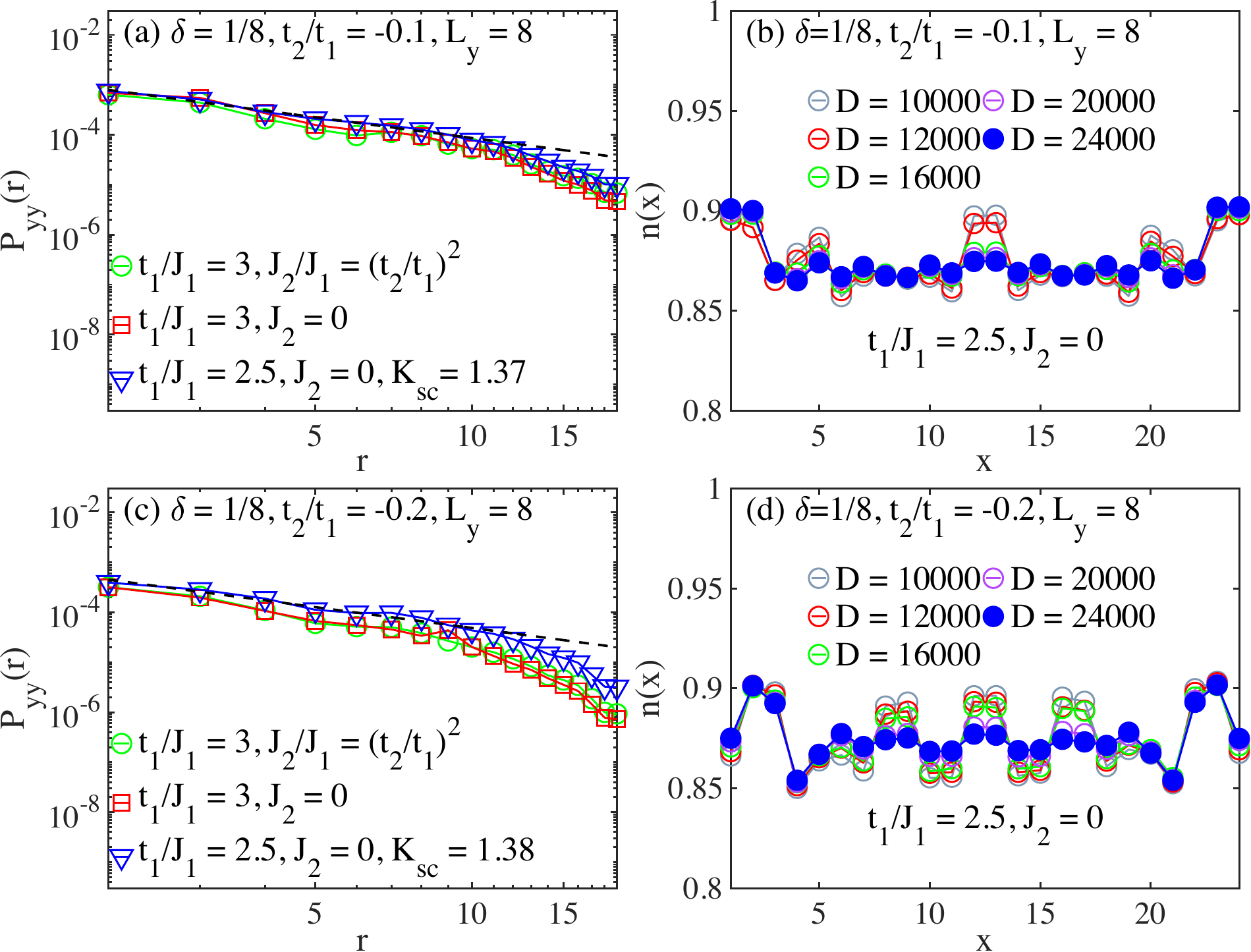}
   \caption{\label{Com_SC2}
   Robust SC states at different model parameters on the eight-leg cylinders at $\delta = 1/8$. (a) Pairing correlations $P_{yy}(r)$ for $t_1/J_1 = 3$, $J_2/J_1 = (t_2/t_1 )^2$; $t_1/J_1 = 3$, $J_{2}=0$; and $t_1/J_1 = 2.5$, $J_2 = 0$. We fix $t_2/t_1=-0.1$. The data are those extrapolated to $D\rightarrow \infty$. The algebraic fitting of the results at $t_1/J_1 = 2.5$, $J_2 = 0$ gives $K_{\rm sc} = 1.37$. (b) Charge density distributions $n(x)$ obtained under different bond dimensions for $t_2/t_1 = -0.1$, $t_1/J_1 = 2.5$, $J_2 = 0$. (c) and (d) are similar plots for $t_2/t_1 = -0.2$.}
\end{figure}

{\it Robust SC phases at different model parameters.---}
In the study of extended $t$-$J$ models, the $t_1$-$t_2$-$J_1$ model with $t_1/J_1 = 2.5$ has also been widely considered~\cite{Corboz_PRL_2014,White_PNAS_2021,Tohyama_PRB_2018}.
To confirm the discovered SC phases at hole doping for different model parameters, we further examine the $t_1$-$t_2$-$J_1$ model with $t_1/J_1 = 2.5$ and $3.0$ ($J_2=0$).
By comparing the pairing correlation and charge density distribution on six- and eight-leg systems (see Fig.~\ref{Com_SC2} and SM~\cite{SM}), we confirm that the identified SC phases are robust against both a small change of $t_1/J_1$ and the absence of $J_2$ interaction.

{\it Summary and discussion.---}
We have presented a global picture for both the electron-doped ($t_2>0$) and hole-doped ($t_2<0$) $t$-$J$ models by DMRG calculations.
While we confirm the $d$-wave SC for electron doping~\cite{White_PNAS_2021} on wider cylinders, we find that the ground states of the hole-doped case can also be superconducting, at both the low doping regime $\delta=1/36-1/24$ for $L_y=6$ and optimal doping $\delta=1/8$ for $L_y=8$ 
with $d$-wave symmetry. 
For $\delta = 1/8$ at hole doping, SC turns out to be favored on wider system, where the enhanced phase coherence of paired holes~\cite{lu2023sign} helps to destabilize CDW and thus allows superconductivity to develop.

Despite the strong competition between stripe and SC orders under hole doping~\cite{White_PNAS_2021}, the SC phases we obtain on both six- and eight-leg systems are stable against a small tuning of $t_1/J_1$, and therefore are established as a common phase for different extended $t$-$J$ models.
Thus, we conclude that the single-band $t$-$J$ model has some generic features including the uniform SC at electron doping, and the dominant SC with near vanishing or coexisting CDW order at hole doping, which may provide a basic description of the cuprate superconductors.

Finally, we discuss some open questions. 
For the hole-doped $t$-$J$ model, the charge order with suppressed SC is commonly observed  as the ground states of narrower systems ($L_y = 6$) with hole binding~\cite{White_PNAS_2021,lu2023sign}, which may have some connection with the pseudogap physics~\cite{lee2014,dai2020} of cuprate systems. 
The $d$-wave SC on the electron-doped side turns out to be robust on wider cylinders. However the nature of its magnetic order is still under debate~\cite{Gong_PRL_2021,White_PNAS_2021}. 
While our analyses of spin correlation lengths suggest a magnetic order at small doping  $\delta \simeq 1/24$, we find the magnetic order is  suppressed for $\delta = 1/8$ as the ratio of 
$\xi_{\rm s} / L_y$ reduces with increased $L_y$~\cite{SM}. 
For the stripe phase at $L_y = 8$ (see SM~\cite{SM}), the CDW order appears to be stable with improved bond dimension, but the pairing correlations keep growing slowly, showing a possible tendency to develop a weak quasi-long-range SC. 
We believe our work will stimulate more future studies to address these challenging issues.

{\it Acknowledgments.---}
We thank Zheng-Yu Weng, Steven Kivelson, Hongchen Jiang, Shengtao Jiang, and Steven White for stimulating discussions. X.~L. and S.~S.~G. were supported by the National Natural Science Foundation of China (Grants No. 12274014 and No. 11834014). W.~Z. was supported by National R\&D program under No. 2022YFA140220, R\&D Program of Zhejiang (2022SDXHDX0005). F.~C. and D.~N.~S. was supported by the U.S. Department of Energy, Office of Basic Energy Sciences under Grant No. DE-FG02-06ER46305 for studying unconventional superconductivity. S. S. G. also acknowledges the support from the Dongguan Key Laboratory of Artificial Intelligence Design for Advanced Materials.

{\it Note added.---}
At the final stage of preparing this work, we have become aware of an independent and related work focusing on the larger positive $t_{2}/t_1 \simeq 0.7$ regime of the same $t$-$J$ model on eight-leg cylinder~\cite{Jiang_Kivelson_Lee_2023}, as well as two other works focusing on the Hubbard model~\cite{xu2023coexistence, Jiang2023t}.
The results in Ref.~\cite{Jiang_Kivelson_Lee_2023} are consistent with our findings at $t_2/t_1=0.3$.

\bibliography{refs}

\clearpage

\appendix
\widetext
\begin{center}
	\textbf{\large Supplemental Materials for: “Emergent Superconductivity and Competing Charge Orders in Hole-Doped Square-Lattice $t$-$J$ Model”}
\end{center}

\vspace{1mm}

\renewcommand\thefigure{\thesection S\arabic{figure}}
\renewcommand\theequation{\thesection S\arabic{equation}}

\setcounter{figure}{0} 
\setcounter{equation}{0} 

In the Supplemental Materials, we provide more numerical results to support the conclusions we have discussed in the main text. In Sec. A, we show the bond dimension scaling of correlation functions in the different phases, and we analyze their decay exponents and correlation lengths.
In Sec. B, we supplement the singlet pairing correlations for more parameter points and the triplet pairing correlations on the eight-leg systems at $\delta = 1/8$ doping level.
In Sec. C, we present the charge density profiles and bond energy distributions in the different phases. In particular, we show how the charge density oscillation is suppressed with growing bond dimension in the superconducting (SC) phases on the eight-leg systems. 
In Sec. D, we demonstrate the static spin structure factor in the different phases and the spin correlation lengths for different system circumferences at lower doping levels.
In Sec. E, we show the electron momentum distribution $n(\mathbf{k})$ in the different phases.
In Sec. F, we discuss the physical properties of the $\mathrm{W_{y}3}$ charge density wave (CDW) phase.
In Sec. G, we demonstrate the robust SC phases at different $t$-$J$ model parameters (including the $t_1$-$t_2$-$J_1$ model), for both the six- and eight-leg systems.
In Sec. H, we present the details of the extrapolation of correlation functions with increasing bond dimension.

\section{\label{sec:Scaling}A. Correlation functions on the six- and eight-leg cylinders}

In this section, we further discuss the numerical results of various correlation functions on the six- and eight-leg cylinders to provide more supports for the data shown in the main text. 

\begin{figure}
   \includegraphics[width=1.0\textwidth,angle=0]{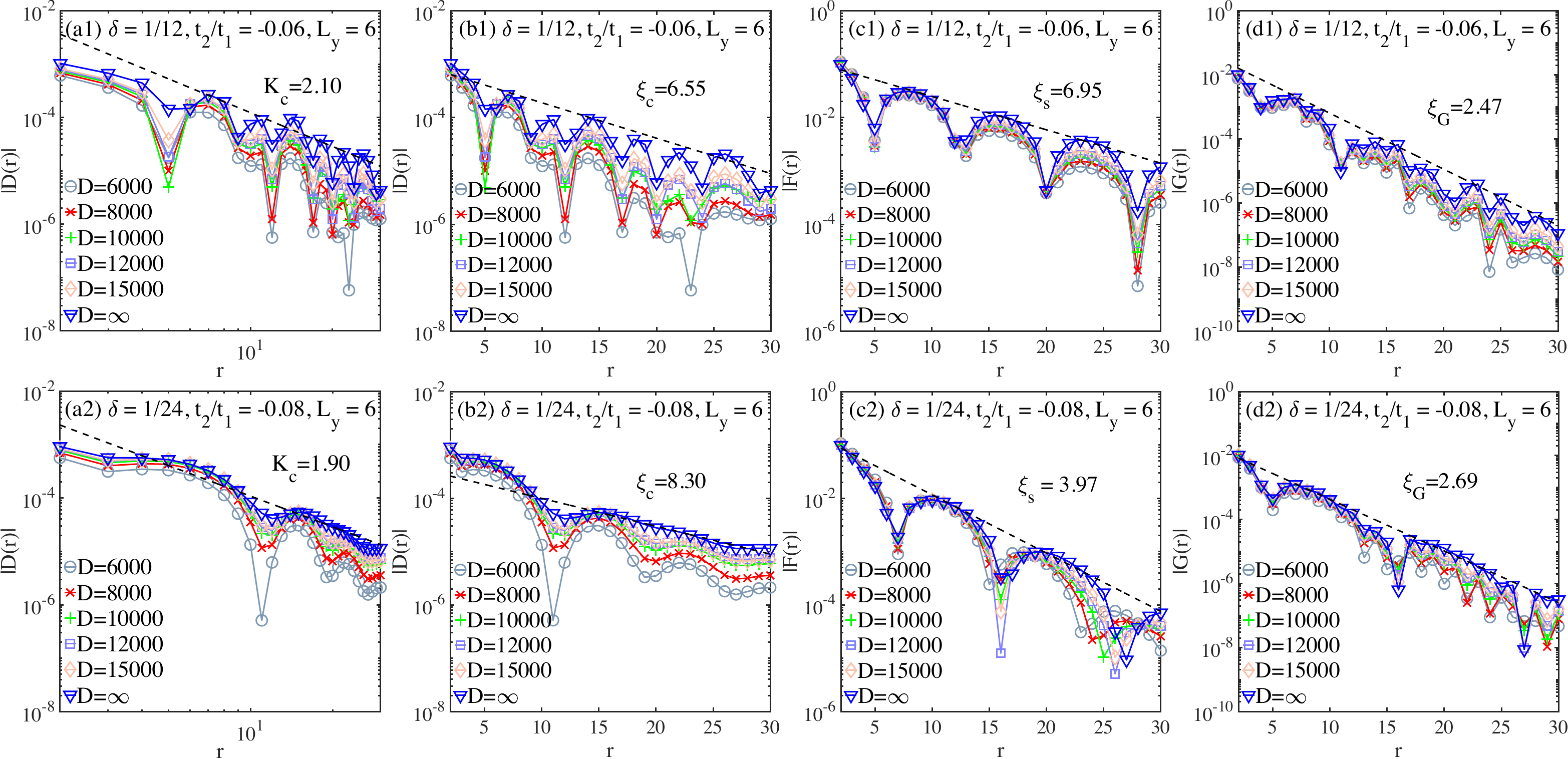}
   \caption{\label{Scaling_correlations}
   Scaling of the correlations in the stripe phase and SC + CDW phase on the $L_y = 6, L_x = 48$ cylinders.
   (a1) and (b1) are respectively the double- and semilogarithmic plots of charge density correlation $D(r)$ obtained by different bond dimensions in the stripe phase. (c1) and (d1) are the semilogarithmic plots of spin correlation $F(r)$ and single-particle Green’s function $G(r)$, respectively. The dashed lines denote the fitting of the extrapolated $D\to \infty$ results, where the power exponents and correlation lengths are obtained by algebraic and exponential fitting, respectively. (a2-d2) are the similar results in the SC + CDW phase.}
\end{figure}

As shown in Figs.~\ref{Scaling_correlations}(a1)-\ref{Scaling_correlations}(d1), we perform the polynomial bond dimension extrapolation (see the details in Sec. H) for the charge density correlation $D(r)$, spin correlation $F(r)$, and single-particle Green’s function $G(r)$ in the stripe phase on the six-leg cylinder. 
We keep the bond dimensions up to $15000$ $\rm SU(2)$ multiplets (equivalent to about $45000$ U(1) states), which lead to a good convergence for our results. 
For density correlation $D(r)$, we present two ways of plotting as shown in Figs.~\ref{Scaling_correlations}(a1) and \ref{Scaling_correlations}(b1), where the power exponent $K_{\mathrm{c}}$ is slightly larger than $2$ in doublelogarithmic scale, while it also can be fitted exponentially with $\xi_{\mathrm{c}} \simeq 6.55$. 
For $F(r)$ [Fig.~\ref{Scaling_correlations}(c1)] and $G(r)$ [Fig.~\ref{Scaling_correlations}(d1)], one can find that these two correlations well follow the exponential decay $\mathrm{exp}\left(-r/\xi \right)$, giving the corresponding correlation lengths $\xi_{\mathrm{s}} \simeq 6.95$ and $\xi_{\mathrm{G}} \simeq 2.47$. 

Following the same procedure, we extrapolate the various correlation functions in the SC + CDW phase on the six-leg cylinder as shown in Figs.~\ref{Scaling_correlations}(a2)-\ref{Scaling_correlations}(d2). 
The density correlation in the SC + CDW phase also can be fitted algebraically with $K_{\mathrm{c}}\simeq1.90$ as shown in Fig.~\ref{Scaling_correlations}(a2), and one can find that the related exponents may not follow the general behavior of the Luther-Emery liquid, i.e. $K_{\mathrm{sc}}=1/K_{\mathrm{c}}$. These results may indicate a crossover of the quasi-long ranged SC order from dominated by the strong quantum fluctuations of one-dimensional system to be more like a two-dimensional (finite size) system.
We also try to plot $D(r)$ in semilogarithmic scale as shown in Fig.~\ref{Scaling_correlations}(b2), which could yield a relatively larger correlation length with $\xi_{\mathrm{c}}\simeq8.30$. 
This density correlation has also been compared with the pairing correlation as shown in Fig.~\ref{Correlations_Stripe_SC}(b), showing that the density correlation is highly intertwined with the pairing correlation. Hence, we identify the phase where the SC and CDW orders are coexistent as shown in Fig.~\ref{Geo-Pha2}(a).
One can see that both the spin correlation in Fig.~\ref{Scaling_correlations}(c2) and single-particle Green’s function in Fig.~\ref{Scaling_correlations}(d2) are short ranged with $\xi_{\mathrm{s}} \simeq 3.97$ and $\xi_{\mathrm{G}} \simeq 2.69$ respectively.

\begin{figure}
   \includegraphics[width=1.0\textwidth,angle=0]{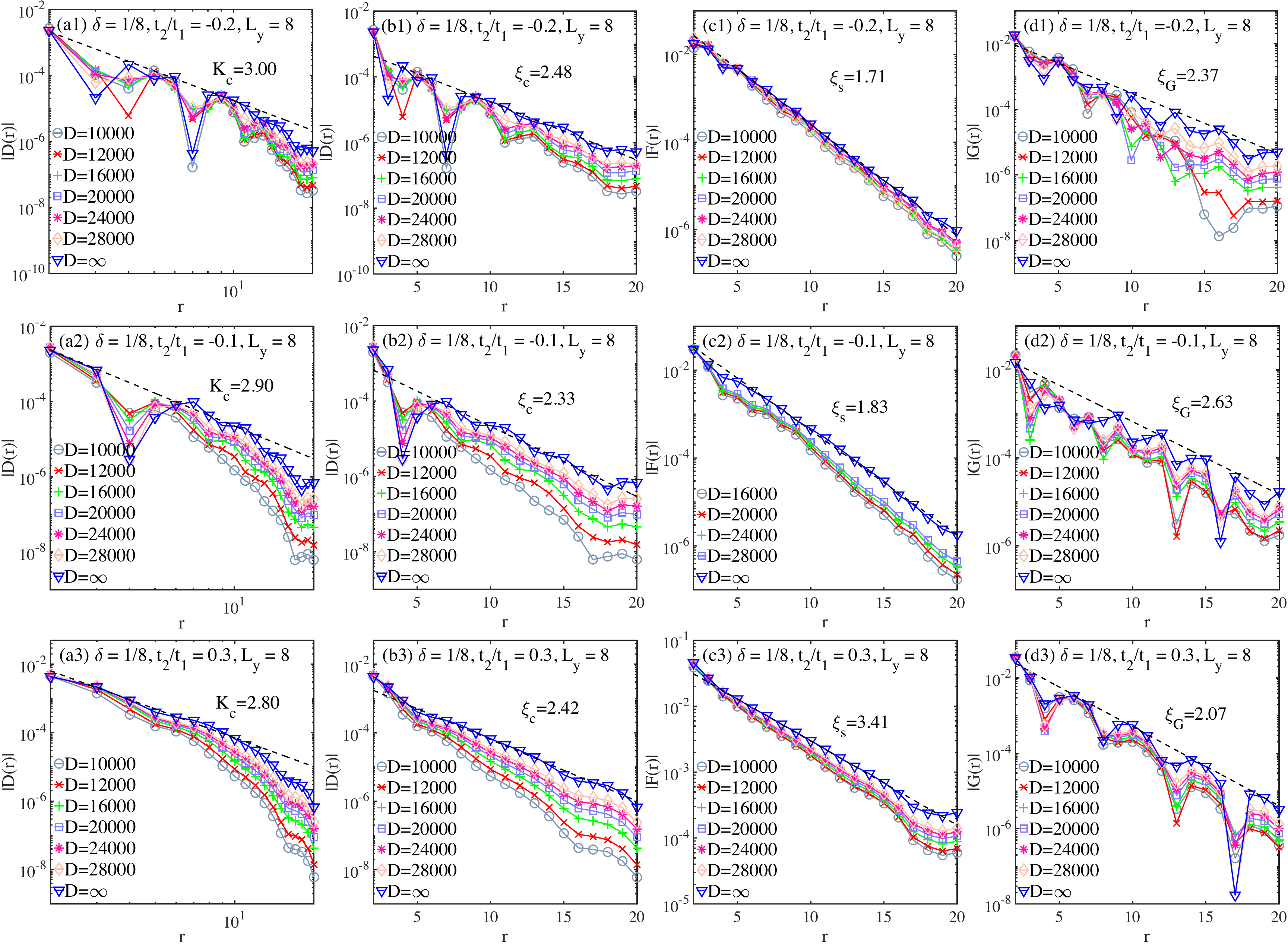}
   \caption{\label{Scaling_correlations_8}
   Scaling of the correlations in the SC and uniform $d$-wave SC phase on the $L_y = 8, L_x = 24$ cylinders.
    (a1) and (b1) are respectively the double- and semilogarithmic plots of charge density correlation $D(r)$ obtained by different bond dimensions in the SC phase ($\delta=1/8, t_2/t_1=-0.2$). (c1) and (d1) are the semilogarithmic plots of spin correlation $F(r)$ and single-particle Green’s function $G(r)$ in the same phase separately. The dashed line denotes the fitting of the extrapolated $D\to \infty$ results, where the corresponding power exponent $K_{\mathrm{c}}$ and correlation length $\xi$ are obtained by fitting the algebraic behaviour $D(r) \sim r^{-K_{\mathrm{c}} }$ and exponential behaviour $\sim \mathrm{exp}\left(-r/\xi \right)$. (a2-d2) and (a3-d3) are respectively the similar correlations in the SC phase ($\delta=1/8, t_2/t_1=-0.1$) and uniform $d$-wave SC phase ($\delta=1/8, t_2/t_1=0.3$), respectively.}
\end{figure}

By the same way, we extrapolate these correlation functions on the eight-leg cylinder as shown in Fig.~\ref{Scaling_correlations_8}. Since the computational complexity increases exponentially with the increase of $L_y$, we keep the bond dimensions up to $28000$ $\rm SU(2)$ multiplets for a good convergence, which are equivalent to about $84000$ U(1) states and reach our simulation limit.
One can see that in both the hole-doped SC and the uniform $d$-wave SC phases, the correlation functions $D(r)$, $F(r)$, and $G(r)$ could be well characterized by exponential decay. 
Although we can always plot the charge density correlations in doublelogarithmic scale as shown in Figs.~\ref{Scaling_correlations_8}(a1), \ref{Scaling_correlations_8}(a2) and \ref{Scaling_correlations_8}(a3), their fitting power exponents are clearly larger than $2$. Hence, considering the nearly uniform charge density distributions, the density correlation results suggest no quasi-long-range charge order~\cite{Arrigoni_PRB_2004}. 
We also find that $K_{\mathrm{c}} K_{\mathrm{sc}} = 1$ may be not well satisfied in the two phases.

For the optimal $1/8$ doping, we systematically compare the correlation functions at different system circumferences.
The obtained power exponents and correlation lengths at different $t_2 / t_1$ ($t_2/t_1 = -0.2 - 0.3$) are summarized in Table~\ref{Table I}.
In the SC and SC + CDW phases, we demonstrate the power exponent $K_{\rm sc}$ for pairing correlation.
For obtaining the charge order exponent $K_{\rm c}$, we fit the charge density correlation $D(r)$.
In some systems, the fittings of the short-distance and the longer-distance density correlation data give slightly different exponents $K_{\rm c}$, both of which are shown in Table~\ref{Table I}.
For spin correlation and single-particle Green's function, we present the correlation lengths obtained from exponential fitting.

In the main text, we mention that for the $L_y = 8$ stripe phase, the pairing correlations slowly increase with bond dimension (see Fig.~\ref{Correlation_SC_8}(b) below), showing a possible tendency to develop a quasi-long-range SC order at $D \rightarrow \infty$.
Therefore, in Table~\ref{Table I} we present both the power exponent $K_{\rm sc}$ and correlation length $\xi_{\rm sc}$ for the $L_y = 8$ systems at $t_2/t_1 = 0, 0.1$.
One can see that the algebraic fitting of the shorter-distance pairing correlations can give the power exponents $K_{\rm sc}$ only slightly larger than $2$.
While our current accessible bond dimensions cannot pin down the SC in the stripe phase, the exact nature of this regime deserves more studies in future.

In Table~\ref{Table I}, we also present the results of $\xi_{\rm s} / L_y$, which can give an estimation of the magnetic order towards 2D limit.
At $t_2 < 0$, both $\xi_{\rm s}$ and $\xi_{\rm s} / L_y$ are significantly reduced on wider system size, indicating the absent magnetic order.
On the other hand, at $t_2 > 0$, while $\xi_{\rm s}$ increases slightly from $L_y = 6$ to $L_y = 8$, $\xi_{\rm s} / L_y$ has a small decrease. 
Therefore, our current results could be consistent with the absent or very weak magnetic order for $\delta = 1/8$ doping at $t_2 > 0$, which call for future studies on larger system sizes.

Finally, we would like to point out a difference between Ref.~\cite{Jiang_PRR_2020} and our results for $L_y = 4$ system.
In Ref.~\cite{Jiang_PRR_2020}, there are two small phase separation regimes at a moderate $t_2 > 0$ (for example $\delta = 1/8$, $t_2/t_1 = 0.1$), which are described by the existence of different CDW patterns in different parts of the system.
We reexamine these parameters and find that by choosing the proper length $L_x$ we can obtain a certain CDW pattern with a well defined CDW wavelength.
Taking $\delta = 1/8$, $t_2/t_1 = 0.1$ as an example.
While the phase separation state reported in Ref.~\cite{Jiang_PRR_2020} exhibits the half-filled stripe (the wave vector is $4\pi \delta$) and filled stripe (the wave vector is $2\pi \delta$) in different parts of the lattice, we find that if $L_x$ is chosen properly, the obtained ground state has a well defined wave vector $Q$ satisfying $1/Q = 1/(2\pi \delta) + 1/(4\pi \delta)$.

\begin{table*}
   \caption{Summary of the power exponents and correlation lengths for the pairing correlation $P_{yy}(r)$, charge density correlation $D(r)$, single-particle Green's function $G(r)$, and spin correlation $F(r)$ on the $4 \times 48$, $6 \times 32$, and $8 \times 24$ $t$-$J$ cylinders with different $t_2/t_1$ at the fixed doping level $\delta=1/8$. The power exponents from algebraic fitting are denoted as $K$, and the correlation lengths from exponential fitting are denoted as $\xi$. The correlation functions for fitting are obtained from the extrapolation to the infinite-bond-dimension limit. For the density correlations $D(r)$ in some cases, the fittings of the short-distance and the longer-distance data give slightly different exponents $K_{\rm c}$, both of which are shown in the table.}
     \begin{ruledtabular}\label{Table I}
        \begin{tabular}{c c c c c c c c c}
        $L_y \times L_x$ & $t_2/t_1$ & $\delta$ &  Phase & $K_{\mathrm{sc}}$ & $K_{\mathrm{c}}$ & $\xi_{\mathrm{G}}$ & $\xi_{\mathrm{s}}$ & $\xi_{\mathrm{s}}/L_y$ \\
        \hline
                    & $-0.2$ & $1/8$ & SC + CDW & 1.32 & 0.77 & 2.22 & 4.82 & 1.21 \\
                    & $-0.1$ & $1/8$ & SC + CDW & 1.23 & 0.79 & 2.18 & 4.71 & 1.18 \\
   $4 \times 48$        & $0$    & $1/8$ & SC + CDW & 1.13 & 0.95 & 2.10 & 4.23 & 1.06 \\
                    & $0.1$  & $1/8$ & SC + CDW & 0.74 & 1.29 & 2.91 & 3.56 & 0.89 \\
                    & $0.2$  & $1/8$ & SC + CDW & 0.84 & 1.23 & 12.23 & 4.39 & 1.10 \\
                    & $0.3$  & $1/8$ & SC + CDW  & 0.95 &  $1.16 - 1.62$ & 7.28 & 4.62 & 1.16 \\ 
        \hline                                    
                    & $-0.2$ & $1/8$ & Stripe & $\xi_{\mathrm{sc}} \approx 2.34$ & 2.35 & 3.98 & 2.20 & 0.37 \\
                    & $-0.1$ & $1/8$ & Stripe & $\xi_{\mathrm{sc}} \approx 2.40$ & 2.26 & 2.32 & 5.94 & 0.99 \\
   $6 \times 32$        & $0$    & $1/8$ & Stripe & $\xi_{\mathrm{sc}} \approx 3.40$ & 1.57 & 2.27 & 5.84 & 0.97 \\
                    & $0.1$  & $1/8$ & SC     & 0.59 & $2.22 - 2.47$ & 2.50 & 2.94 & 0.49  \\
                    & $0.2$  & $1/8$ & SC     & 0.54 & $1.89 - 2.07$ & 2.52 & 2.65 & 0.44 \\
                    & $0.3$  & $1/8$ & SC     & 0.53 & $1.70 - 2.19$ & 2.74 & 2.89 & 0.48  \\ 
        \hline                                    
                    & $-0.2$ & $1/8$ & SC       & 1.63 & $3.00 - 3.30$ & 2.37 & 1.71 & 0.21\\
                    & $-0.1$ & $1/8$ & SC       & 1.46 & $2.90 - 3.12$ & 2.63 & 1.83 & 0.23\\
           & $0$    & $1/8$ & Stripe   & 2.03, $\xi_{\mathrm{sc}} \approx 3.24$ & $2.75 - 3.11$ & 2.69 & 2.19 & 0.27\\
   $8 \times 24$                & $0.1$  & $1/8$ & Stripe   & 2.09, $\xi_{\mathrm{sc}} \approx 2.54$& $2.65 - 3.14$ & 2.01 & 4.02 & 0.50\\
                    & $0.2$  & $1/8$ & SC       & 0.84 & $2.98 - 3.08$ & 1.83 & 2.85 & 0.36\\
                    & $0.3$  & $1/8$ & SC       & 0.57 & $2.80 - 3.02$ & 2.07 & 3.41 & 0.43\\                               
         \end{tabular}     
      \end{ruledtabular}
\end{table*}
\section{\label{sec:Correlations_36}B. Pairing correlation functions for more parameter points}

\begin{figure}
   \includegraphics[width=0.65\textwidth,angle=0]{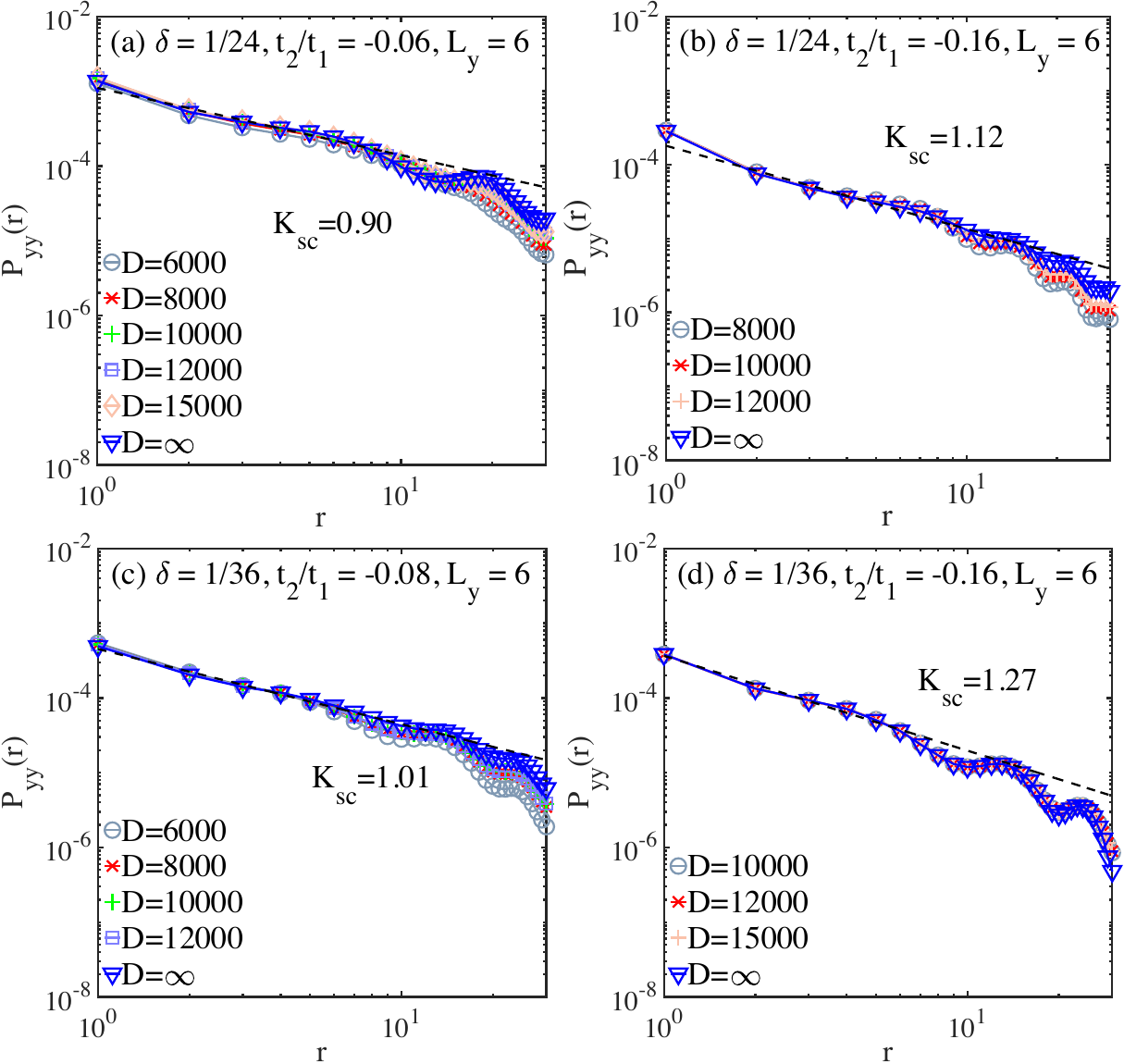}
   \caption{\label{Correlation_SC_36}
   Scaling of pairing correlations for more parameter points in the SC + CDW phase on six-leg cylinder. (a)-(b) doublelogarithmic plots of $P_{yy}(r)$ at $\delta=1/24$ with $t_2/t_1=-0.06$ and $-0.16$. The power exponents $K_{\mathrm{sc}}\simeq 0.90$ and $K_{\mathrm{sc}}\simeq 1.12$ are obtained by fitting $P_{yy}(r) \sim r^{-K_{\mathrm{sc}} }$. The dashed lines denote the fitting of the extrapolated $D\to \infty$ results. (c) and (d) are the similar plots of $P_{yy}(r)$ at $\delta=1/36$ with $t_2/t_1=-0.08$ and $-0.16$. The power exponents are obtained as $K_{\mathrm{sc}}\simeq 1.01$ and $K_{\mathrm{sc}}\simeq 1.27$ respectively.}
\end{figure}

\begin{figure}
   \includegraphics[width=0.95\textwidth,angle=0]{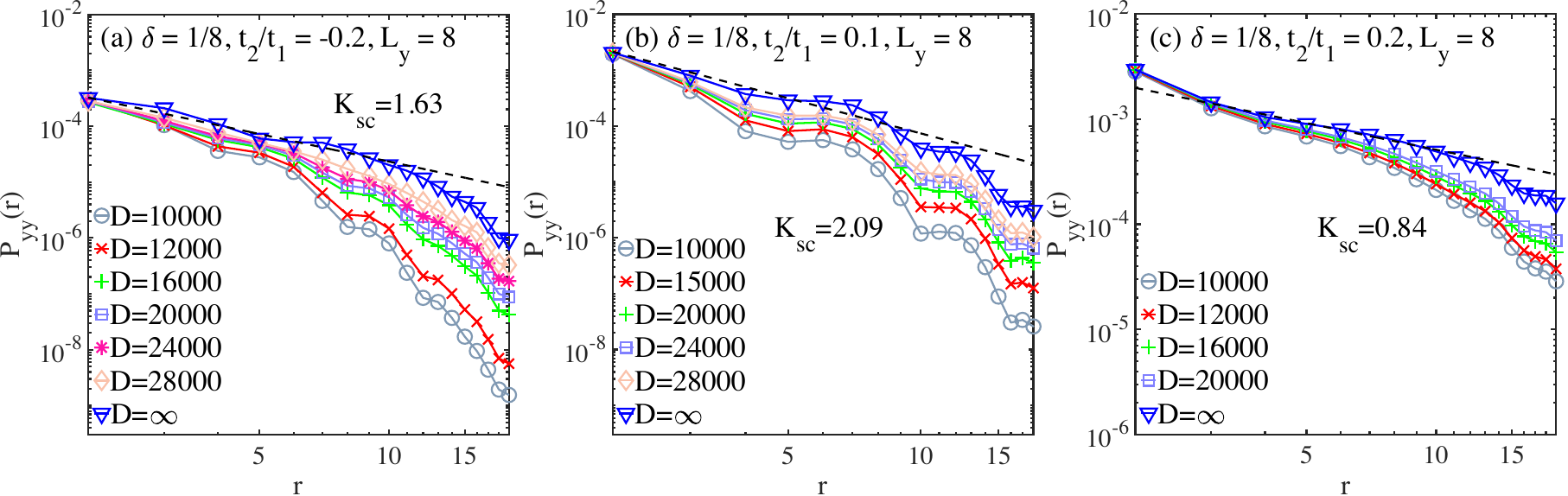}
   \caption{\label{Correlation_SC_8}
   Scaling of pairing correlations for more parameter points on eight-leg cylinder at $\delta = 1/8$ doping level. (a) and (c) are the plots of $P_{yy}(r)$ at $t_2/t_1=-0.2$ and $0.2$, for the hole-doped and electron-doped SC phases, respectively. The power exponents are obtained as $K_{\mathrm{sc}}\simeq 1.63$ and $K_{\mathrm{sc}}\simeq 0.84$ by fitting $P_{yy}(r) \sim r^{-K_{\mathrm{sc}} }$. The dashed lines denote the fitting of the extrapolated $D\to \infty$ results. (b) is the similar plot of $P_{yy}(r)$ at $t_2/t_1=0.1$ in the stripe phase, where the power exponent is obtained as $K_{\mathrm{sc}}\simeq 2.09$.}
\end{figure}

Here we show the pairing correlations for more parameter points to support the SC phases demonstrated in the phase diagram. 
We first provide additional representative points in the SC + CDW phase on six-leg cylinder. 
In Figs.~\ref{Correlation_SC_36}(a)-\ref{Correlation_SC_36}(b), one can see that the pairing correlations $P_{yy}(r)$ for $\delta=1/24, t_2/t_1=-0.06$ and $-0.16$ can be fitted algebraically with $K_{\mathrm{sc}} \simeq 0.90$ and $K_{\mathrm{sc}} \simeq 1.12$ after bond dimension extrapolation. 
The consistent results can be also found at $\delta=1/36, t_2/t_1=-0.08$ and $-0.16$ [Figs.~\ref{Correlation_SC_36}(c)-\ref{Correlation_SC_36}(d)], where the power exponents are $K_{\mathrm{sc}} \simeq 1.01$ and $K_{\mathrm{sc}} \simeq 1.27$, respectively. 
For the hole-doped SC phase on the eight-leg cylinder, we also show the scaling of pairing correlations at $\delta=1/8, t_2/t_1=-0.2$ [Fig.~\ref{Correlation_SC_8}(a)], where the extrapolated $P_{yy}(r)$ can also be fitted algebraically with $K_{\mathrm{sc}} \simeq 1.63$. 
For $\delta=1/8, t_2/t_1=0.2$ in the electron-doped SC phase [Fig.~\ref{Correlation_SC_8}(c)], the power exponent is obtained as $K_{\mathrm{sc}} \simeq 0.84$. 
These results further support the SC phases presented in the main text.
For the stripe phase at $\delta = 1/8$ doping on eight-leg cylinder, we have also analyzed the pairing correlations.
Up to the bond dimensions $D = 28000$, we find that the pairing correlations keep increasing slowly as shown in Fig.~\ref{Correlation_SC_8}(b) at $t_2/t_1 = 0.1$.
After the bond dimension scaling and by fitting the short-distance data, we obtain a power exponent $K_{\mathrm{sc}} \simeq 2.09$, which is close to $2$ and indicates a possible quasi-long-range SC order that however cannot be determined within our accessible bond dimensions.
On the other hand, since the CDW pattern appears robust in this regime, as shown in Fig.~\ref{CDW_8}(b), we denote it as a stripe phase.

\begin{figure}
   \includegraphics[width=0.7\textwidth,angle=0]{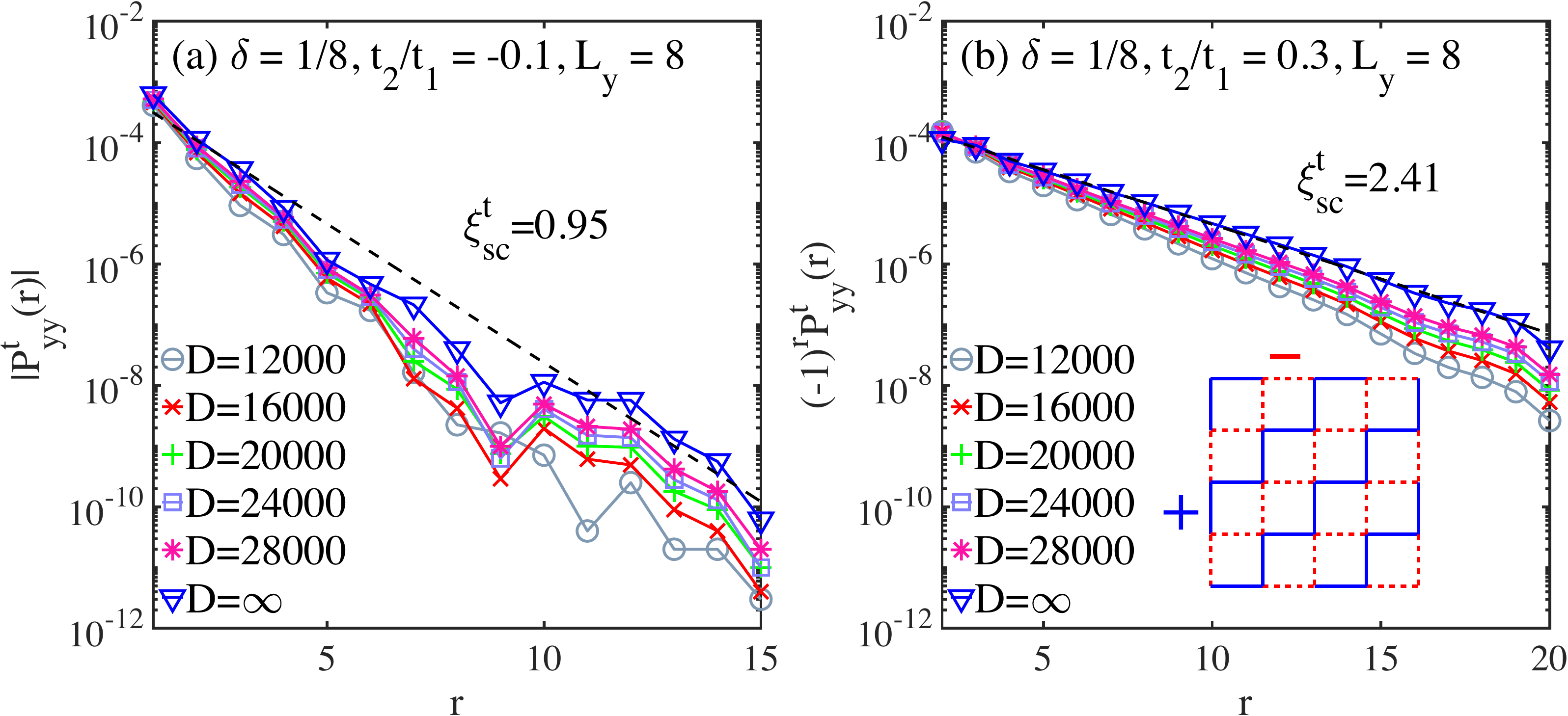}
   \caption{\label{TripletSC}
   Triplet SC pairing correlations $P^{t}_{yy}(r)$ on the eight-leg cylinder at $\delta=1/8$. (a) semilogarithmic plot of $|P^{t}_{yy}(r)|$ for $t_2/t_1=-0.1$ in the hole-doped SC phase. The dashed line denotes the exponential fitting of the extrapolated $D\to \infty$ results, giving a very short correlation length $\xi^{t}_{\mathrm{sc}}\simeq 0.95$. (b) semilogarithmic plot of $(-1)^{r} P^{t}_{yy}(r)$ for $t_2/t_1=0.3$ in the uniform $d$-wave SC phase. The triplet pairing correlation exhibits a good exponential decay, giving a short correlation length $\xi^{t}_{\mathrm{sc}}\simeq 2.41$. The signs of the triplet pairing correlations between different bonds agree with the $(\pi,\pi)$ $p$-wave symmetry as shown by the inset.}
\end{figure}

In Ref.~\cite{White_PNAS_2021}, a coexistence of $d$-wave SC and triplet $(\pi,\pi)$ $p$-wave SC is reported on the eight-leg $t_1$-$t_2$-$J_1$ model at $t_2/t_1 > 0$.
To examine a possible triplet SC in our study, we also measure the triplet pairing correlations $P^{t}_{\alpha, \beta }(\mathbf{r}) = \langle {\hat{\Delta}}_{\alpha,t }^{\dagger } ({\mathbf{r}}_0) {\hat{\Delta}}_{\beta,t }({\mathbf{r}}_0 +\mathbf{r}) \rangle$, where the triplet pairing order parameter is defined as ${\hat{\Delta} }_{\alpha,t } \left(\mathbf{r}\right)=\left({\hat{c} }_{\mathbf{r}\uparrow } {\hat{c} }_{\mathbf{r}+{\mathbf{e}}_{\alpha } \downarrow } +{\hat{c} }_{\mathbf{r}\downarrow } {\hat{c} }_{\mathbf{r}+{\mathbf{e}}_{\alpha } \uparrow } \right)/\sqrt{2}$ with ${\mathit{\mathbf{e}}}_{\alpha = x,y}$ denoting the unit vectors along the $x$ and $y$ directions. 
We check our model on the eight-leg systems at $\delta = 1/8$ doping level.
On the hole-doped side at $t_2/t_1 = -0.1$ [Fig.~\ref{TripletSC}(a)], $|P^{t}_{yy}(r)|$ decays very fast without a clear pairing symmetry, and the correlation length $\xi^{t}_{\mathrm{sc}}$ is smaller than $1$. 
On the electron-doped side at $t_2/t_1 = 0.3$ [Fig.~\ref{TripletSC}(b)], the triplet pairing correlations between different bonds agree with the $(\pi,\pi)$ $p$-wave symmetry reported in Ref.~\cite{White_PNAS_2021}, but the triplet pairing correlations decay exponentially with a short correlation length $\xi^{t}_{\mathrm{sc}} \simeq 2.41$, which indicates the vanished triplet SC and only a quasi-long-range singlet $d$-wave SC order.

\section{\label{sec:CDW}C. Charge density profiles and bond energy distribution}

\begin{figure}[h]
\includegraphics[width=0.8\textwidth,angle=0]{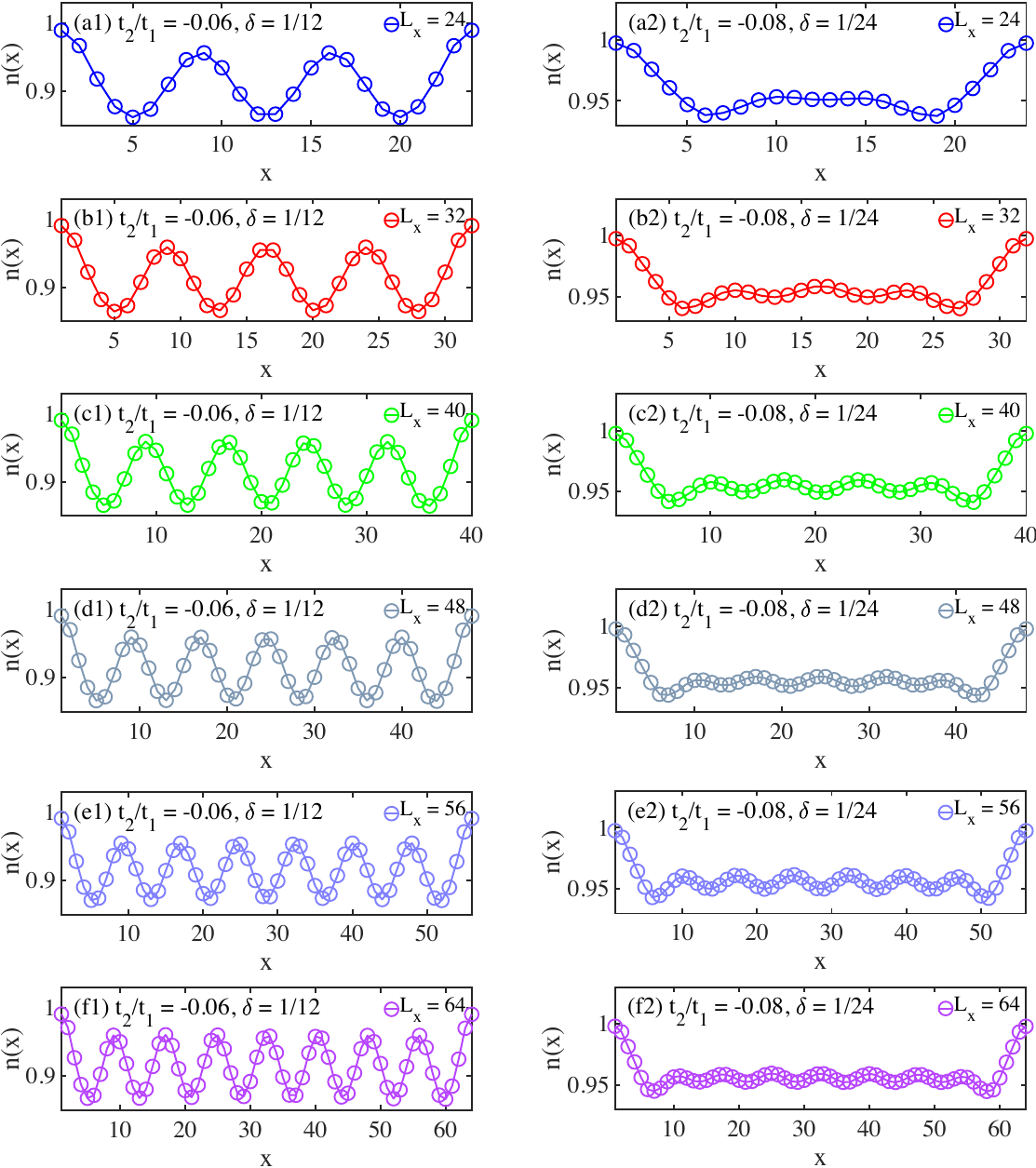}
   \caption{\label{CDW_period}
   Charge density profiles on the six-leg cylinders. Charge density profiles with different lattice sizes $L_{x}=24-64$ in the stripe phase (a1-f1) and SC + CDW phase (a2-f2). Here, we select two representative parameters to show the charge density distributions of these two phases. The truncation error of all simulations are about $10^{-6}$.}
\end{figure}

In the main text, we have shown the charge density profiles in the stripe and SC + CDW phases on the six-leg systems.  
To exclude the finite-size effect, we further examine the charge density distributions on different cylinder lengths ($L_{x}=24-64$) by fixing the doping ratio, as shown in Fig.~\ref{CDW_period}.
Following the definition in the main text, the average charge density in each column is defined as $n(x)=\frac{1}{L_y }\sum_{y=1}^{L_y } \left\langle {\hat{n} }_{x,y} \right\rangle$, where ${\hat{n} }_{x,y} =\sum_{\sigma } {\hat{c} }_{(x,y),\sigma }^{\dagger } {\hat{c} }_{(x,y),\sigma }$ is the electron number operator defined on the site $(x,y)$. 
In Figs.~\ref{CDW_period}(a1)-\ref{CDW_period}(f1), we show the charge density profiles of $t_{2}/t_{1}=-0.06$ and $\delta=1/12$ in the stripe phase. 
One can see that the wavelength for different $L_x$ is always $\lambda\simeq4/(L_{y}\delta)=8$, i.e. $Q=(3\pi\delta,0)$, which remains true for other doping levels in the stripe phase. That is to say, each stripe is filled with four holes, consistent with the CDW phase at the $t_{2}/t_{1}>0$ side~\cite{Gong_PRL_2021}. 
Similarly, we also show the charge density profiles in the SC + CDW phase as shown in Figs.~\ref{CDW_period}(a2)-\ref{CDW_period}(f2) for $t_{2}/t_{1}=-0.08$ and $\delta=1/24$. 
Different from the stripe phase, the wavelength in the SC + CDW phase is $\lambda\simeq2/(L_{y}\delta)=8$, i.e. $Q=(6\pi\delta,0)$.
Such a pattern of two holes filled in each stripe has also been observed in the SC + CDW phase on the six-leg cylinder systems with $t_{2}/t_{1}>0$ ~\cite{Gong_PRL_2021}. 

\begin{figure}[h]
   \includegraphics[width=0.85\textwidth,angle=0]{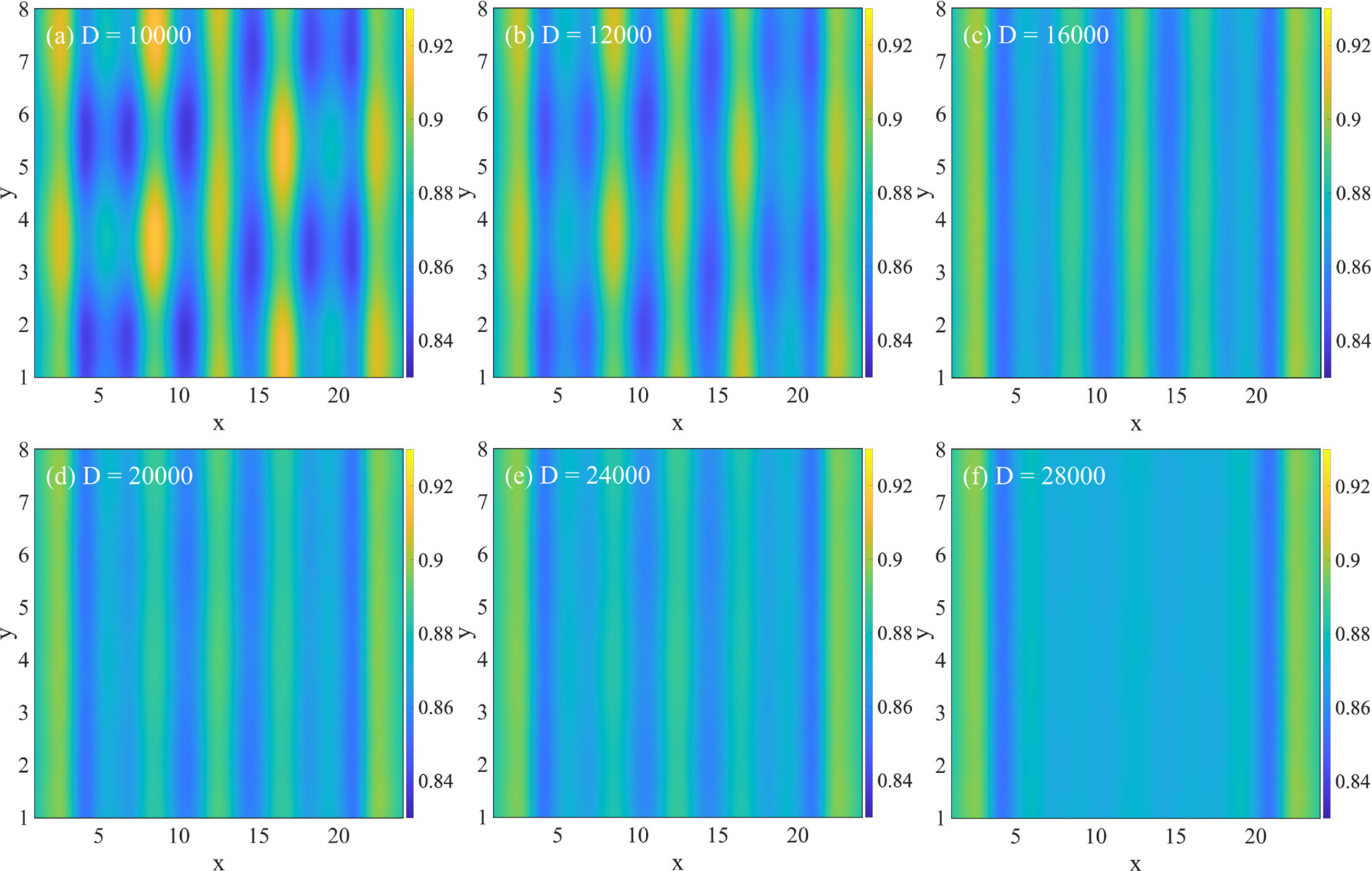}
   \caption{\label{CDW_2D_8}
   Bond dimension dependence of the charge densities of all the sites in the hole-doped SC phase on the eight-leg cylinder at $\delta = 1/8$ doping level. Here, we show the results at $t_{2}/t_{1}=-0.2$ on the $L_y = 8, L_x = 24$ cylinder. $x$ and $y$ denote the numbers of column and row, respectively. The color shows the charge density at each site. (a)-(f) show the charge density distributions for $D=10000$, $D=12000$, $D=16000$, $D=20000$, $D=24000$ and $D=28000$, respectively.}
\end{figure}

In the main text, we have also shown the charge density profiles for $t_2/t_1 = -0.2$ in the hole-doped SC phase on the eight-leg cylinder, which have a quite weak charge density oscillation. 
Here we supplement the evolution of the charge densities for all the sites with growing bond dimension, as shown in Fig.~\ref{CDW_2D_8}. One can see that the charge density distributions are even not uniform along the circumference direction at lower bond dimensions, which is repaired with growing bond dimension when $D \geq 16000$ and the charge density oscillation is gradually suppressed.

\begin{figure}
   \includegraphics[width=0.85\textwidth,angle=0]{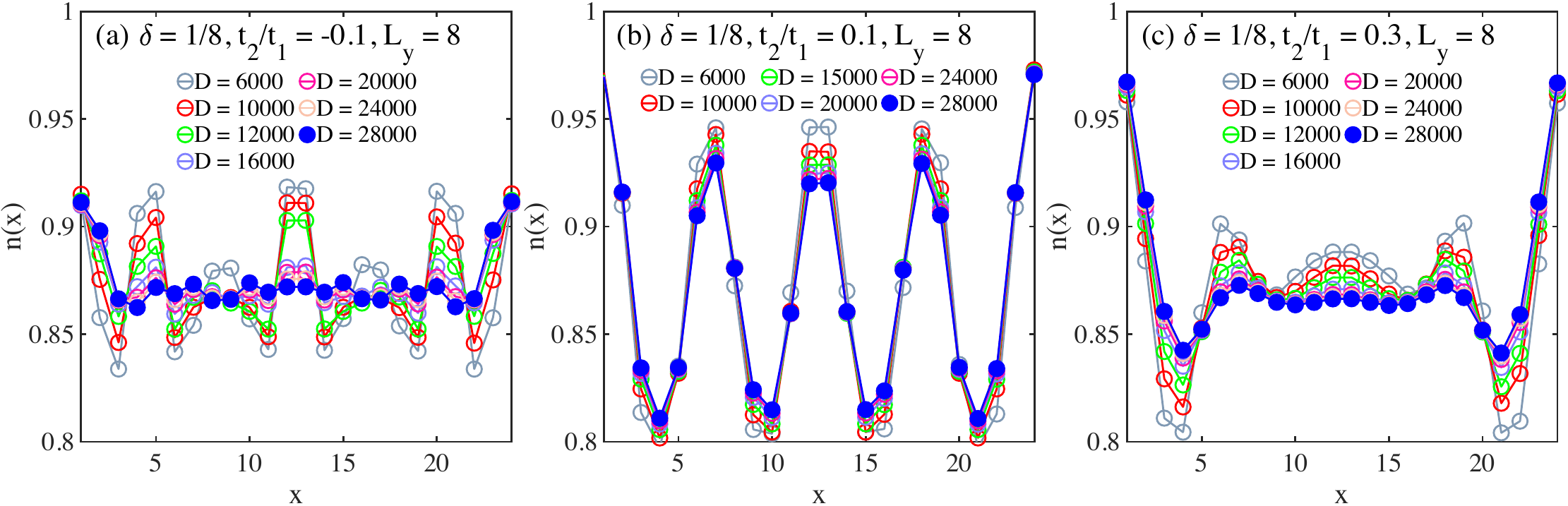}
   \caption{\label{CDW_8}
   Bond dimension dependence of the charge density profiles on the eight-leg systems at $\delta = 1/8$ doping level. (a) and (c) show the results on the $L_y = 8, L_x = 24$ cylinder for the hole-doped SC phase at $t_2/t_1 = -0.1$ and uniform $d$-wave SC phase at $t_2/t_1 = 0.3$, respectively. (b) shows the results on the $L_y = 8, L_x = 24$ cylinder for the stripe phase at $t_2/t_1 = 0.1$.}
\end{figure}

In Fig.~\ref{CDW_8}, we compare the charge density profiles of the different phases on the eight-leg systems at $\delta = 1/8$ doping level.
For the hole-doped SC phase [Fig.~\ref{CDW_8}(a)] and uniform $d$-wave SC phase [Fig.~\ref{CDW_8}(c)], charge density distributions tend to be uniform with increased bond dimensions.
In particular, for the uniform $d$-wave SC phase at  $t_2/t_1 >0$, the uniform charge density distributions persist from six-leg to eight-leg systems~\cite{Gong_PRL_2021}.
In contrast, the CDW pattern appears to be stable with growing bond dimension in the stripe phase [Fig.~\ref{CDW_8}(b)].

\begin{figure}[h]
   \includegraphics[width=0.9\textwidth,angle=0]{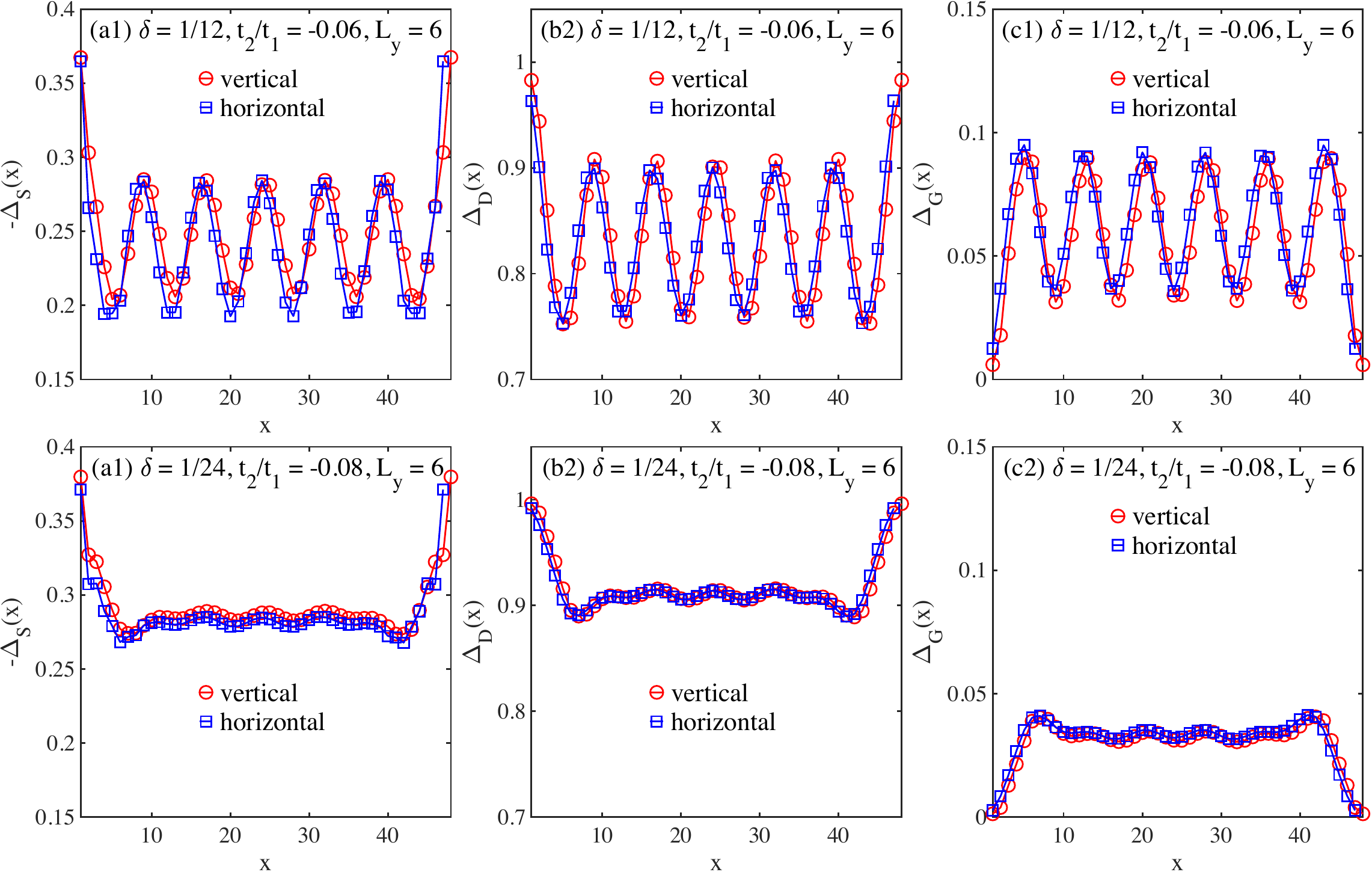}
   \caption{\label{BOW}
   Distributions of the nearest-neighbor energy terms on the six-leg cylinder. (a1-c1) are respectively the spin exchange term $\Delta_{S}=\langle {\hat{\bf S}}_i \cdot {\hat{\bf S}}_j \rangle$, density-density interaction term $\Delta_{D}=\langle \hat{n}_i \hat{n}_j\rangle$, and hopping term $\Delta_{G}=\sum_{\sigma} \langle \hat{c}^{\dagger}_{i,\sigma} \hat{c}_{j,\sigma} + h.c. \rangle$ in the stripe phase. Both of the vertical and horizontal nearest-neighbor energy terms are shown. (a2-c2) are the similar plots in the SC + CDW phase.}
\end{figure}

Furthermore, we analyze the distributions of the nearest-neighbor energy terms in the different phases (examples on the six-leg systems are shown in Fig.~\ref{BOW}). 
We present the nearest-neighbor spin exchange term $\Delta_{S}=\langle {\hat{\bf S}}_i \cdot {\hat{\bf S}}_j \rangle$, density-density interaction term $\Delta_{D}=\langle \hat{n}_i \hat{n}_j\rangle$, and hopping term $\Delta_{G}=\sum_{\sigma} \langle \hat{c}^{\dagger}_{i,\sigma} \hat{c}_{j,\sigma} + h.c. \rangle$ for both vertical and horizontal directions. 
We find that these bond energy terms are always connected to the charge density distributions, without additional bond order emerging.

\section{\label{sec:AF order}D. Static spin structure factor and spin correlation length}

\begin{figure}
   \includegraphics[width=1.0\textwidth,angle=0]{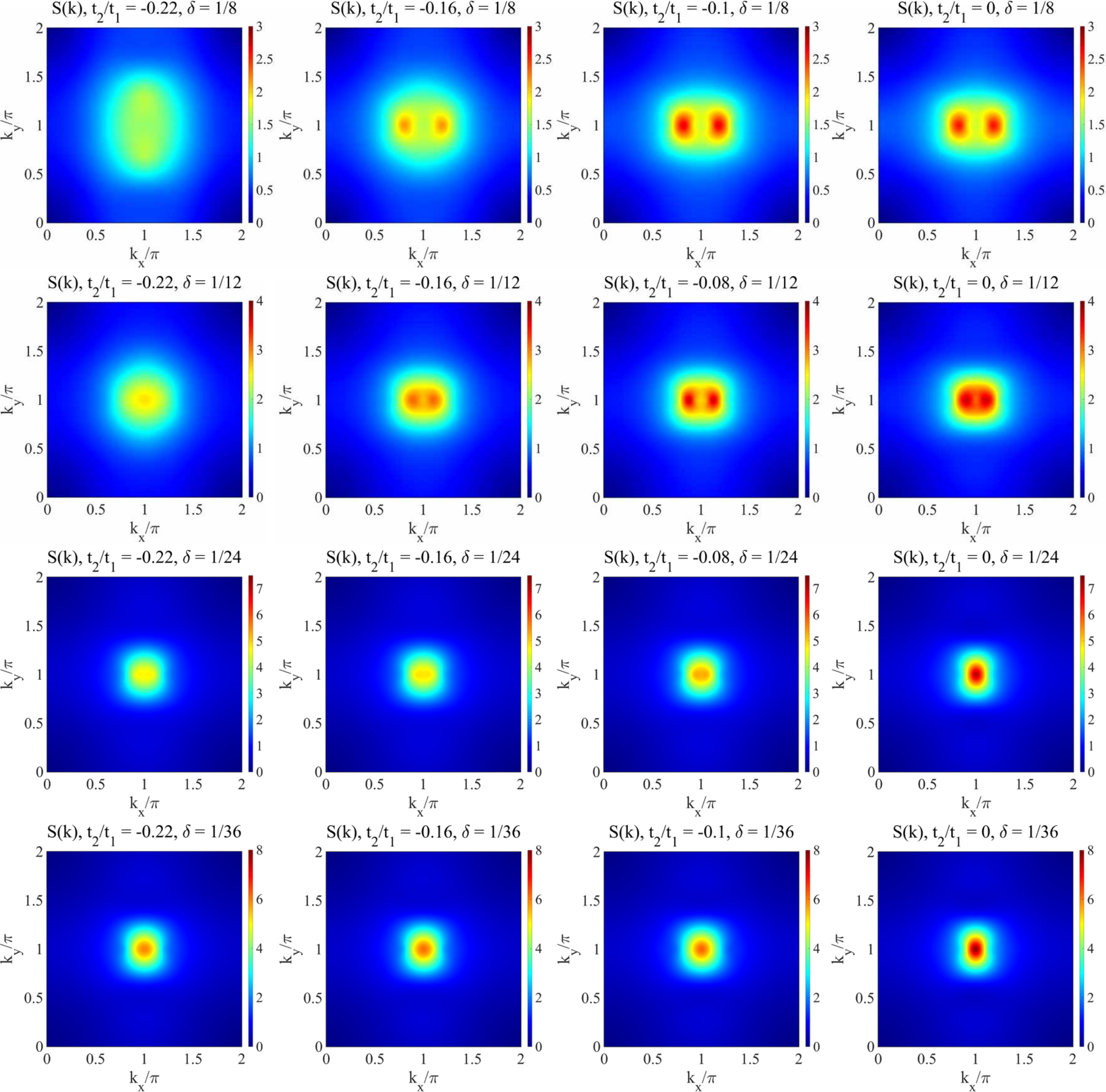}
   \caption{\label{Sk_momentum}
   Static spin structure factors $S(\mathbf{k})$ for different $t_2/t_1$ and doping rations $\delta$ on the six-leg cylinder.
   $S(\mathbf{k})$ are obtained by taking the Fourier transformation for the spin correlations of the middle $32\times 6$ sites on the $48\times 6$ cylinders. All these measurements are obtained by keeping bond dimensions $D=12000$.
}
\end{figure}

\begin{figure}
   \includegraphics[width=1.0\textwidth,angle=0]{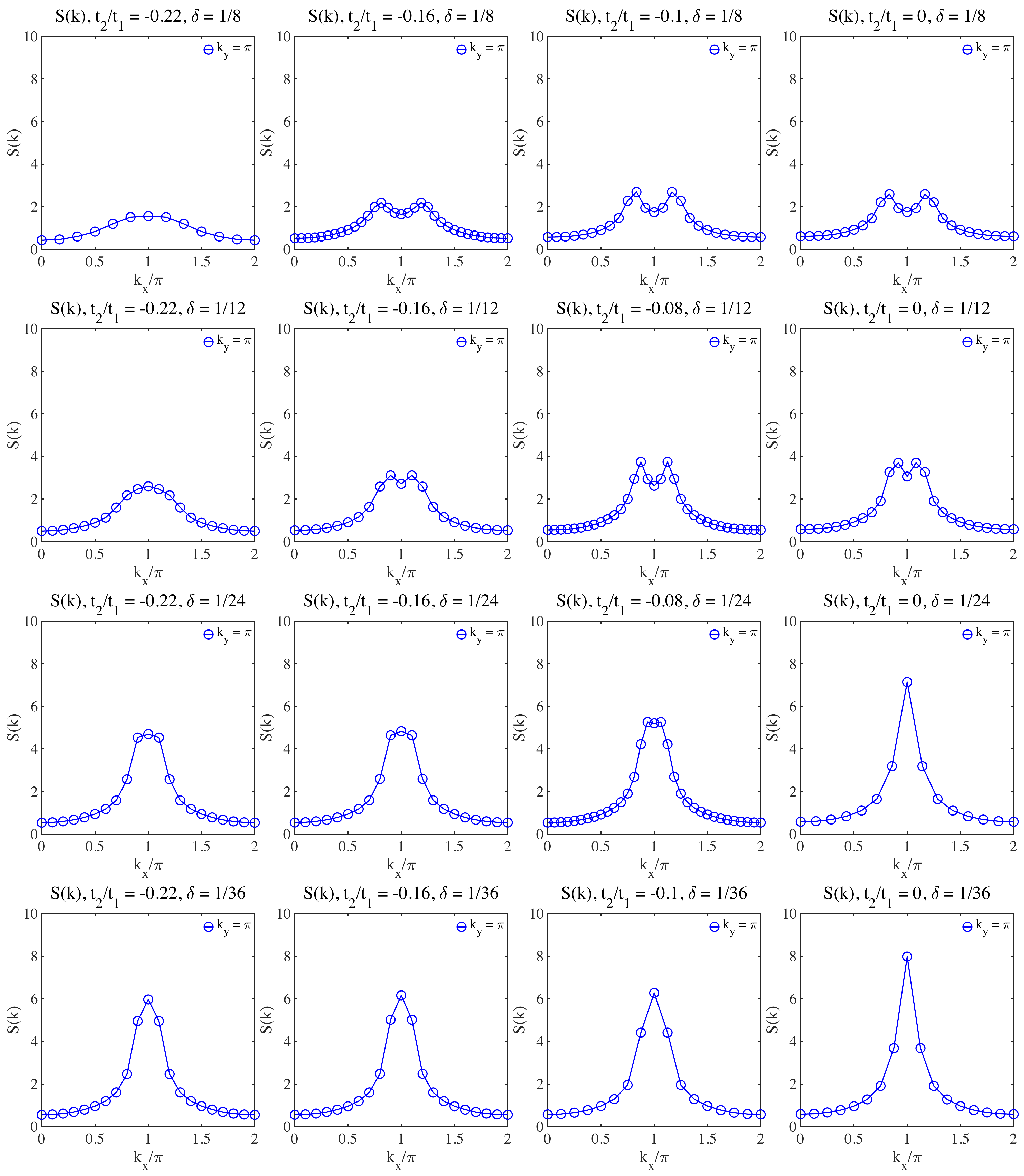}
   \caption{\label{Spin_pi}
   Slice of the static spin structure factors $S(\mathbf{k})$ for different $t_2/t_1$ and doping rations $\delta$ on the six-leg cylinder. The data are the same as those shown in Fig.~\ref{Sk_momentum}. Here the results for $k_{y}=\pi$ are shown.}

\end{figure}

\begin{figure}
   \includegraphics[width=0.7\textwidth,angle=0]{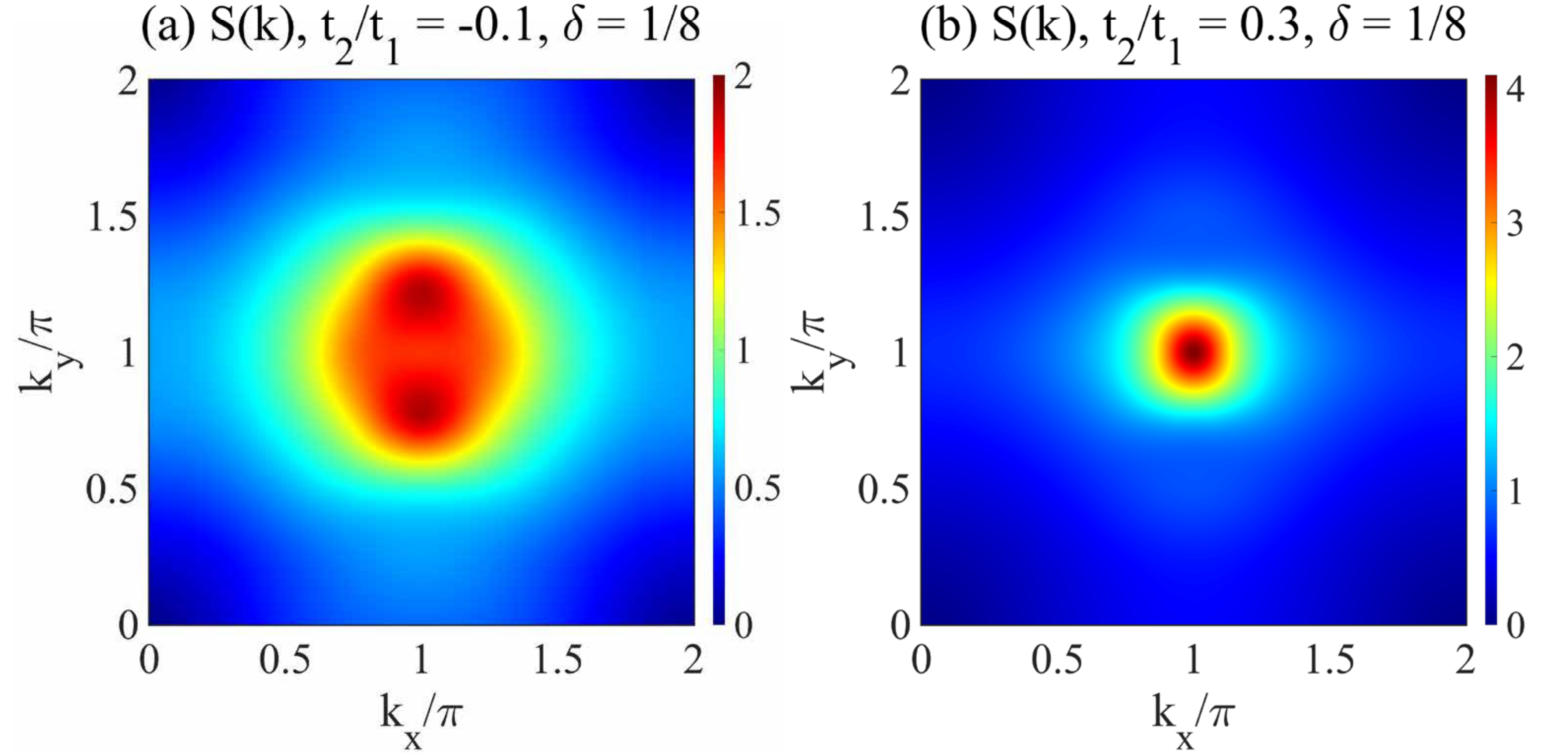}
   \caption{\label{Sk_8}
   Static spin structure factors $S(\mathbf{k})$ in the hole-doped SC phase ($t_2/t_1 = -0.1$) and uniform $d$-wave SC phase ($t_2/t_1 = 0.3$) on the eight-leg cylinder at $\delta = 1/8$ doping level.
   $S(\mathbf{k})$ are obtained by taking the Fourier transformation for the spin correlations of the middle $16 \times 8$ sites on the $24 \times 8$ cylinder. In the hole-doped SC phase (a), there are two round peaks around $\mathbf{k} = (\pi,\pi)$, while 
   a single peak in the uniform $d$-wave SC phase (b). All the measurements are obtained by keeping bond dimensions $D=16000$.
}
\end{figure}

In this section, we demonstrate the static spin structure factor $S(\mathbf{k})$ in different phases and analyze the spin correlation length $\xi_{\rm s}$ at the lower doping levels. 
On the six-leg cylinder [Fig.~\ref{Sk_momentum} and Fig.~\ref{Spin_pi}], the spin correlations generally are stronger at lower doping levels, which agrees with the observation at $t_2/t_1>0$ on the six-leg systems~\cite{Gong_PRL_2021}.
In the stripe phase, $S(\mathbf{k})$ should have two peaks around $\mathbf{k} = (\pi,\pi)$, which characterize the antiferromagnetic spin correlation with a $\pi$-phase shift. 
This feature is clear in the stripe phase at the larger doping levels such as $\delta = 1/8$ and $1/12$. 
For the smaller doping levels $\delta = 1/24, 1/36$, we confirm the antiferromagnetic spin correlation with the $\pi$-phase shift in real space, but since the subsystem length for computing structure factor [$32$ in Fig.~\ref{Sk_momentum}] is relatively short compared with the long CDW wavelength at lower doping, only a single peak at $\mathbf{k} = (\pi,\pi)$ is obtained due to the limit of resolution.
In the SC + CDW phase, spin correlations also have the phase shift [see Fig.~\ref{Scaling_correlations}(c2)] and the same resolution limit of $S(\mathbf{k})$ also exists in Fig.~\ref{Sk_momentum}.
For the eight-leg systems at $\delta = 1/8$ doping level [Fig.~\ref{Sk_8}], $S(\mathbf{k})$ exhibits two broad peaks near $\mathbf{k} = (\pi,\pi)$ in the hole-doped SC phase [Fig.~\ref{Sk_8}(a)] and a single peak at $\mathbf{k} = (\pi,\pi)$ in the uniform $d$-wave SC phase [Fig.~\ref{Sk_8}(b)]. 
The peak magnitude of $S(\mathbf{k})$ in the uniform $d$-wave SC phase is clearly larger than that in the hole-doped SC phase, indicating the stronger spin correlation. 

Furthermore, to investigate possible magnetic order at lower doping levels, we analyze the spin correlation lengths at different system circumferences and doping levels.
We focus on two representative parameter points $t_2/t_1 = -0.1$ and $0.2$, with the doping levels from $\delta = 1/24$ to $\delta = 1/48$ on the $L_y = 4, 6, 8$ cylinders.
The obtained spin correlation lengths $\xi_{\rm s}$ by exponentially fitting spin correlation functions are summarized in Table~\ref{Table III}.
For both $t_2/t_1 = -0.1$ and $0.2$ on any given system circumference, $\xi_{\rm s}$ always increases with reducing doping level.
For the hole doping at $t_2/t_1 = -0.1$ and doping $\delta>1/48$, both correlation lengths $\xi_{\rm s}$ and $\xi_{\rm s} / L_y$ are significantly reduced on the wider $L_y = 8$ cylinder, which suggest the absence of magnetic order in 2D limit. 
However, for the lower doping $\delta=1/48$, we find spin correlations become sensitive to the system length $L_x$. Much wider and longer systems are required to determine the nature of magnetic order.  
We conjecture that for $t_2/t_1 = -0.1$, the system may develop a magnetic order for doping level $\delta \lesssim  1/48$. 
In contrast, $\xi_{\rm s}$ grows almost linearly with circumference for the electron doping at $t_2/t_1 = 0.2$ and $\delta = 1/24, 1/32, 1/36$, which indicates the much stronger spin correlations compared with that in the hole doping side and the highly possible magnetic order in 2D limit.

\begin{table*}
   \caption{Summary of the spin correlation lengths on the four-leg, six-leg, and eight-leg systems at small doping levels with the given $t_2/t_1 = -0.1$ and $0.2$. The spin correlation lengths $\xi_{\rm s}$ are obtained from the exponential fitting of spin correlation functions, which are shown in the third (for $L_y = 4$), fourth (for $L_y = 6$), and fifth (for $L_y = 8$) columns. The bracket after each value of $\xi_{\rm s}$ includes the cylinder length for the corresponding calculation. The spin correlation functions are obtained with the bond dimensions up to $12000$. In the last column, $\xi_{\rm s} / L_y$ for $L_y = 4, 6, 8$ are shown together for a comparison. For these parameters, the charge density distributions have the periods of $1/(2 \delta)$, $1/(3 \delta)$, and $1/(4 \delta)$ for the four-leg, six-leg, and eight-leg systems, respectively.}
     \begin{ruledtabular}\label{Table III}
     \setlength{\tabcolsep}{5mm}{
        \begin{tabular}{c c c c c c}
        $t_2/t_1$ & $\delta$ & $\xi_{\mathrm{s}} (L_{y}=4)$ &$\xi_{\mathrm{s}} (L_{y}=6)$ & $\xi_{\mathrm{s}} (L_{y}=8)$ & $\xi_{\mathrm{s}}/L_y$ $(L_y = 4, 6, 8)$\\
       \hline
     $-0.1$ & $1/24$ & $5.69$ $(L_x = 48)$ & $3.73$ $(L_x = 48)$ & $2.71$ $(L_x = 24)$ & $1.42$, $0.62$, $0.34$\\
     $-0.1$ & $1/32$ & $5.80$ $(L_x = 48)$ & $5.44$ $(L_x = 32)$ & $3.10$ $(L_x = 24)$ & $1.45$, $0.91$, $0.39$\\
     $-0.1$ & $1/36$ & $5.86$ $(L_x = 36)$ & $6.01$ $(L_x = 48)$ & $3.87$ $(L_x = 24)$ & $1.47$, $1.00$, $0.48$\\
     $-0.1$ & $1/48$ & $5.87$ $(L_x = 48)$ & $6.99$ $(L_x = 48)$ & $4.23$ $(L_x = 24)$ &$1.47$, $1.16$, $0.53$\\
        \hline                                    
     $0.2$ & $1/24$ & $4.0$ $(L_x = 48)$ & $7.34$ $(L_x = 48)$ & $10.12$ $(L_x = 24)$  & $1.00$, $1.22$, $1.27$\\
     $0.2$ & $1/32$ & $4.23$ $(L_x = 48)$ & $7.83$ $(L_x = 32)$ & $11.36$ $(L_x = 24)$ & $1.06$, $1.30$, $1.42$\\
     $0.2$ & $1/36$ & $4.29$ $(L_x = 36)$ & $8.26$ $(L_x = 48)$ & $12.15$ $(L_x = 24)$ & $1.07$, $1.38$, $1.52$\\                           
         \end{tabular}}
      \end{ruledtabular}
\end{table*}

\section{\label{sec:Fermi surafce}E. Electron momentum distribution}

\begin{figure}
   \includegraphics[width=1.0\textwidth,angle=0]{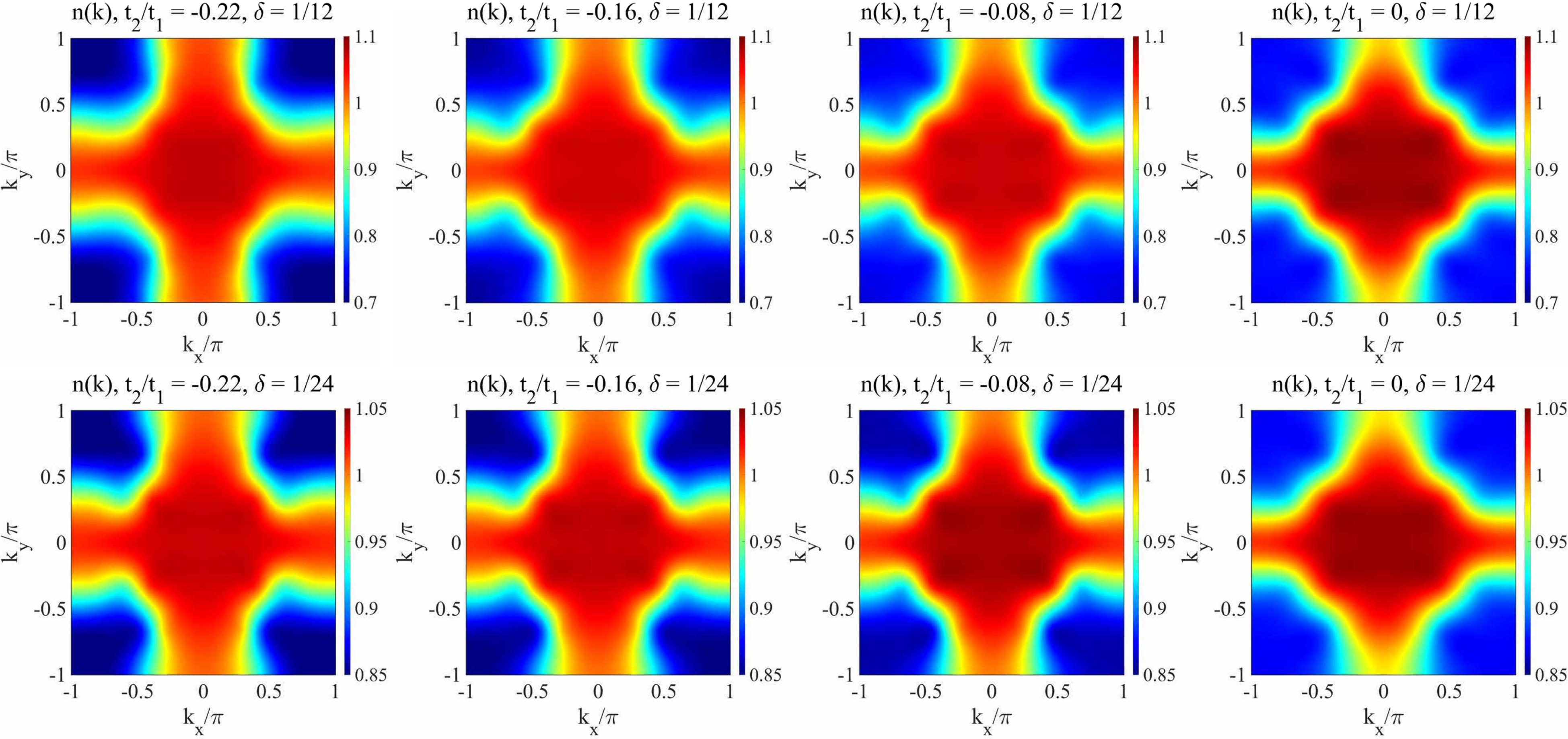}
   \caption{\label{Nk_momentum}
   Momentum distribution functions $n(\mathbf{k})$ for different $t_2/t_1$ and doping rations $\delta$ on the six-leg cylinders.
   $n(\mathbf{k})$ are obtained by taking the Fourier transformation for the single-particle Green's functions of the middle $32 \times 6$ sites on the $48 \times 6$ cylinder. All the measurements are obtained by keeping bond dimensions $D=12000$.
}
\end{figure}

\begin{figure}
   \includegraphics[width=0.7\textwidth,angle=0]{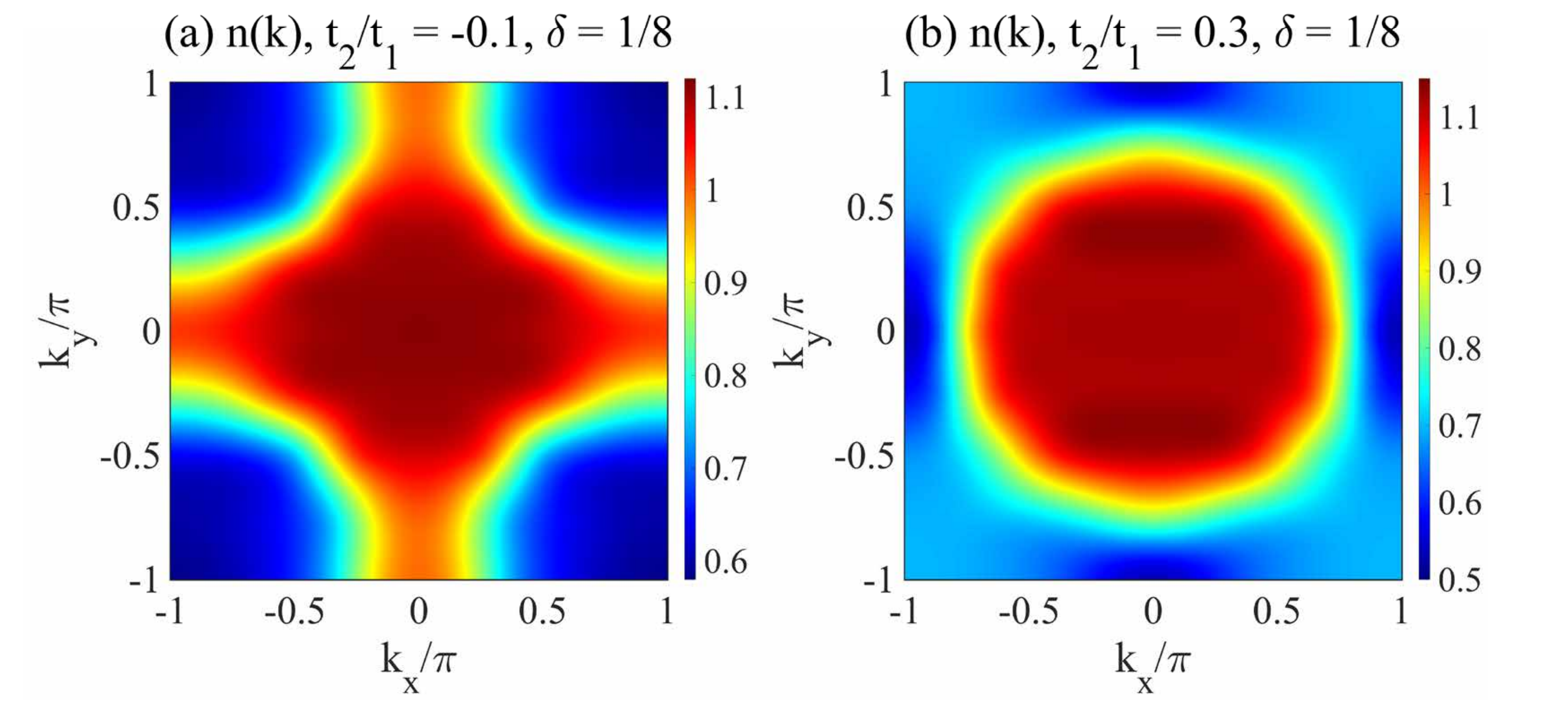}
   \caption{\label{Nk_8}
   Momentum distribution functions $n(\mathbf{k})$ for the hole-doped SC phase ($t_2/t_1 = -0.1$) and uniform $d$-wave SC phase ($t_2/t_1 = 0.3$) on the eight-leg cylinder at $\delta = 1/8$ doping level. $n(\mathbf{k})$ are obtained by taking the Fourier transformation for the single-particle Green's functions of the middle $16 \times 8$ sites on the $24 \times 8$ cylinder. All the measurements are obtained by keeping bond dimensions $D=16000$.
}
\end{figure}

In Fig.~\ref{Nk_momentum}, we provide the electron momentum distribution $n(\mathbf{k})$ for more parameters on the six-leg systems. 
One can find that the topology of the Fermi surface at $\mathbf{k} = \left(\pm \pi ,0\right)$ and $\mathbf{k} = \left(0,\pm \pi \right)$ is unenclosed in all the three phases at $t_2/t_1<0$, which are consistent with the ARPES observation of hole-doped cuprates~\cite{Hossain_nphys_2008}.
The same Fermi surface topology is also found on the eight-leg cylinder at $t_2/t_1 < 0, \delta = 1/8$ [Fig.~\ref{Nk_8}(a)], which is in contrast to the electron-doped case [Fig.~\ref{Nk_8}(b)], where the Fermi surface forms a closed pocket centered around $\mathbf{k}=\left(0,0 \right)$.

\section{\label{sec:Wy3}F. Charge density pattern and correlation functions in the $\mathrm{W_{y}3}$ CDW phase}

\begin{figure}[h]
   \includegraphics[width=0.85\textwidth,angle=0]{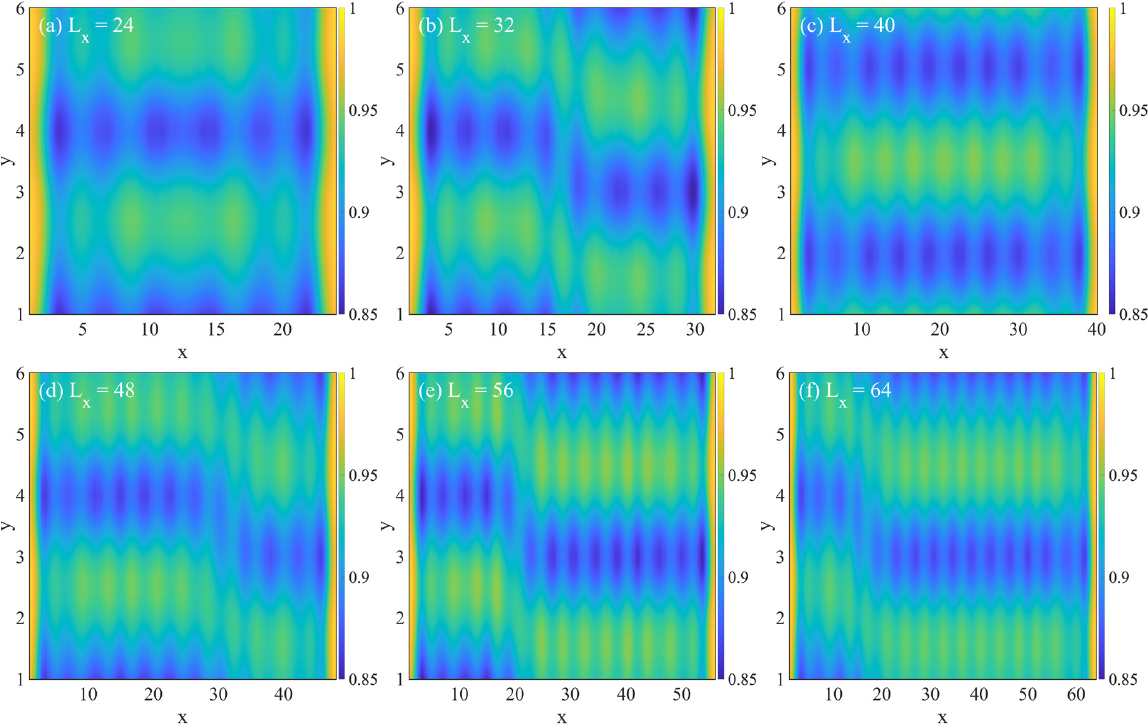}
   \caption{\label{Wy3_stripe}
   Charge density distribution for all the lattice sites in the $\mathrm{W_{y}3}$ CDW phase on the six-leg cylinders. $x$ and $y$ label the column and row numbers, respectively. The color represents the charge density at each site. (a)-(f) are the results for $L_{x}=24$, $L_{x}=32$, $L_{x}=40$, $L_{x}=48$, $L_{x}=56$ and $L_{x}=64$, respectively. Here, we show the results at $\delta = 1/12$, $t_{2}/t_{1}=-0.2$ as a representative. For different $L_x$ at $L_y = 6$, the CDW wavelength along the axis direction is always $4$, satisfying $\lambda \simeq 2/(L_y \delta)$. The truncation error of all the simulations is about $10^{-6}$.}
\end{figure}

\begin{figure}
   \includegraphics[width=0.7\textwidth,angle=0]{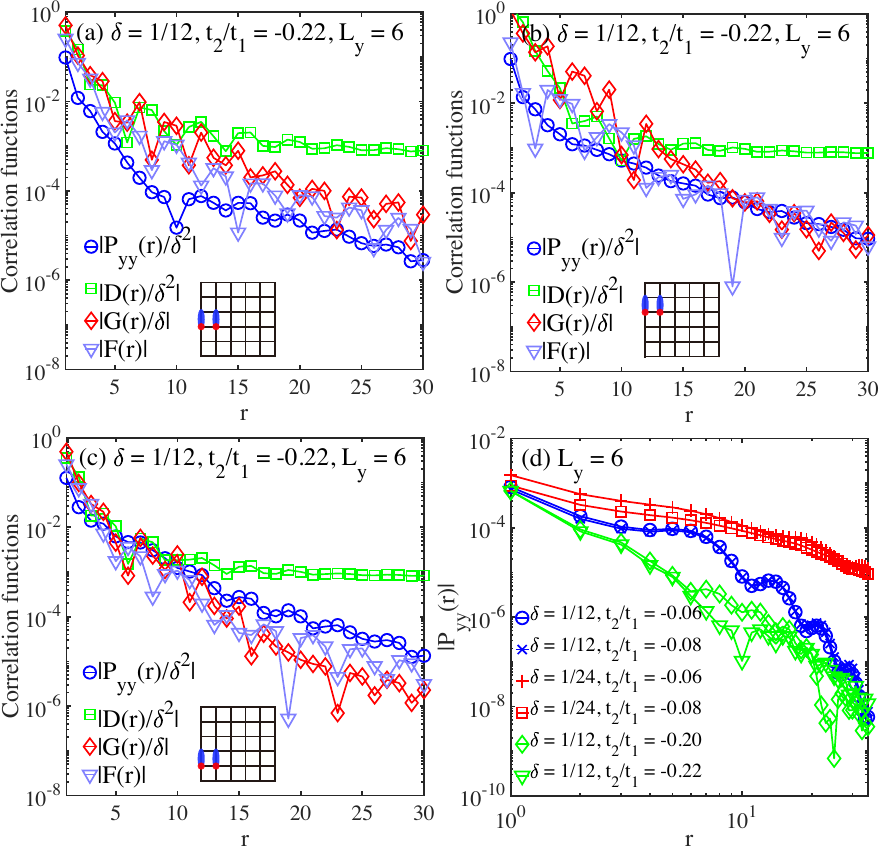}
   \caption{\label{Correlations_Wy3_stripe}
   Correlation functions in the $\mathrm{W_{y}3}$ CDW phase of the six-leg system. (a) Comparison among the pairing correlation $P_{yy}(r)$, charge density correlation $D(r)$, single-particle Green's function $G(r)$, and spin correlation $F(r)$ in the $\mathrm{W_{y}3}$ CDW phase at $t_2/t_1 = -0.22, \delta = 1/12$. The correlations are rescaled by $\delta$ to make a direct comparison. Here, $P_{yy}(r)$ is measured for the vertical bonds as illustrated in the inset. $D(r)$, $G(r)$, and $F(r)$ are measured with the reference site on the third row (the red dots in the inset). (b) and (c) are the similar plots for the other reference bonds and sites, as illustrated by the corresponding insets. (d) doublelogarithmic plot of $P_{yy}(r)$ (following the definition shown in the inset of (a)) for different $t_{2}/t_{1}$ at $\delta=1/12$ and $\delta=1/24$. The blue data for $\delta = 1/12, t_2/t_1 = -0.06, -0.08$ are in the stripe phase. The red data for $\delta = 1/24, t_2/t_1 = -0.06, -0.08$ are in the SC + CDW phase. The greed data for $\delta = 1/12, t_2/t_1 = -0.20, -0.22$ are in the $\mathrm{W_{y}3}$ CDW phase. Here we show the results obtained using the bond dimension $D=15000$ on the $L_x = 48$ cylinder.}
\end{figure}

In this section, we discuss the properties of the $\mathrm{W_{y}3}$ CDW phase. 
As we mentioned in the main text, the charge density distribution of the $\mathrm{W_{y}3}$ CDW phase breaks the translational symmetry in the circumference direction with a period of $3$, and the wavelength of the CDW along the axis direction is $\lambda\simeq2/(L_{y}\delta)$ (without considering the shift of stripes in the $y$ direction). 
To confirm this, we have performed DMRG simulations for different system lengths $L_{x}=24-64$ that can match doping ratio, as shown in Fig.~\ref{Wy3_stripe}. 
One can find that with increase of $L_{x}$, the distributions of charge density are always inhomogeneous in the $y$ direction, suggesting that such an inhomogeneity may not be caused by the finite-size effect. In addition, we have also tried to keep larger bond dimensions and increase the number of sweep, but the translational symmetry breaking in the $y$ direction is not recovered. 
Therefore, the DMRG results suggest this phase as a new kind of CDW phase carrying the wavevector $Q=(6\pi\delta, 2\pi/3)$.

We also find that SC is strongly suppressed in the $\mathrm{W_{y}3}$ CDW phase. 
Since the translational symmetry is broken along the $y$ direction, here we show three kinds of correlations with different reference bonds and sites in Figs.~\ref{Correlations_Wy3_stripe}(a)-\ref{Correlations_Wy3_stripe}(c). 
In Fig.~\ref{Correlations_Wy3_stripe}(a), the middle vertical bonds as illustrated in the inset are selected for exploring pairing correlation $P_{yy}(r)$, and the third row (the numbering is shown in Fig.~\ref{Wy3_stripe}) is selected as the reference sites for exploring other two-site correlations. 
We compare different correlation functions and find that the pairing correlation $P_{yy}(r)$ is the weakest and charge density correlation $D(r)$ is the strongest. 
Similarly, for other cases, while $D(r)$ is still the strongest, $P(r)$, $G(r)$, and $F(r)$ are much weaker. 
In Fig.~\ref{Correlations_Wy3_stripe}(d), we also show the comparison of pairing correlations in different phases on the six-leg cylinder (the vertical bonds are chosen following the inset of Fig.~\ref{Correlations_Wy3_stripe}(a)). 
One can see that $P_{yy}(r)$ are much weaker in the $\mathrm{W_{y}3}$ CDW phase.

\section{\label{sec:benchmark }G. Robust superconducting phases at different model parameters}

\begin{figure}
   \includegraphics[width=0.75\textwidth,angle=0]{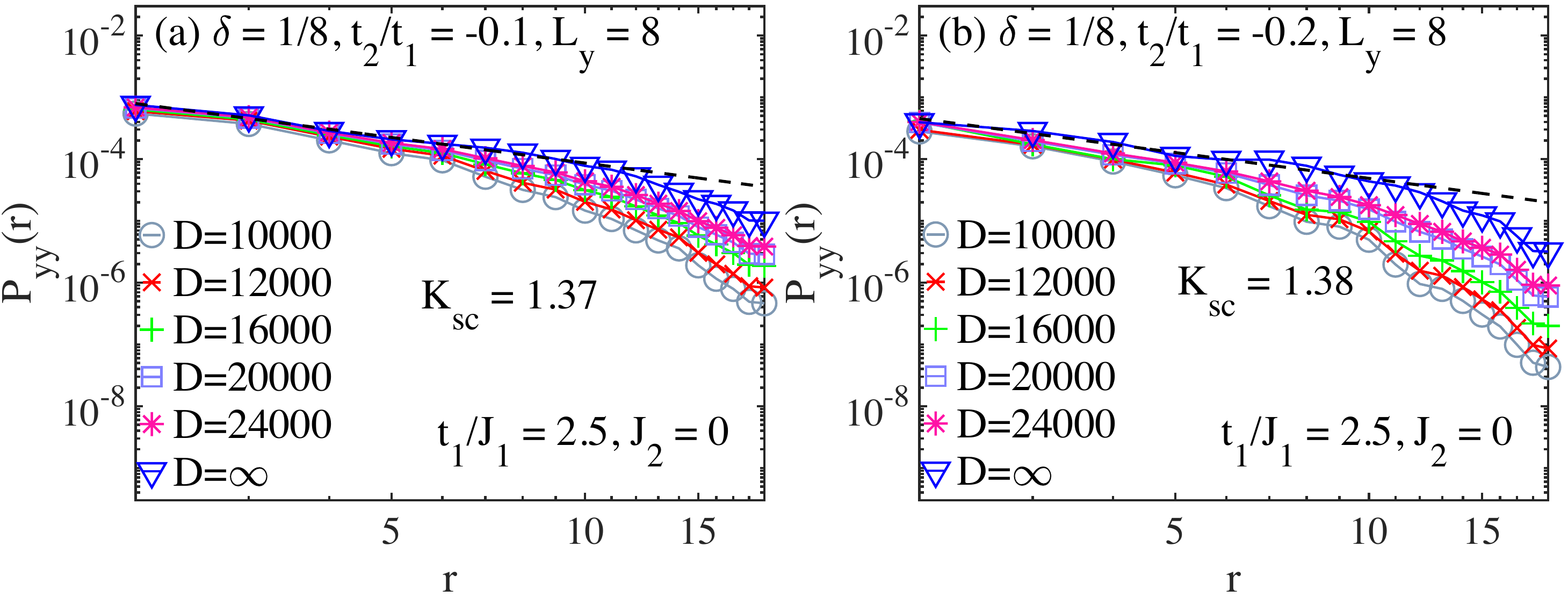}
   \caption{\label{scaling825}
   Bond dimension scaling of the pairing correlation $P_{yy}(r)$ on the eight-leg $t_1$-$t_2$-$J_1$ model with $t_1/J_1= 2.5, J_{2}=0, \delta = 1/8$. (a) doublelogarithmic plot of the pairing correlation $P_{yy}(r)$ for $t_2/t_1= -0.1$. The fitted power exponent for the bond dimension extrapolated results is $K_{\rm sc} \simeq 1.37$. (b) The similar plot of $P_{yy}(r)$ for $t_2/t_1= -0.2$, where  $K_{\rm sc} \simeq 1.38$.}
\end{figure}

\begin{figure}
   \includegraphics[width=1.0\textwidth,angle=0]{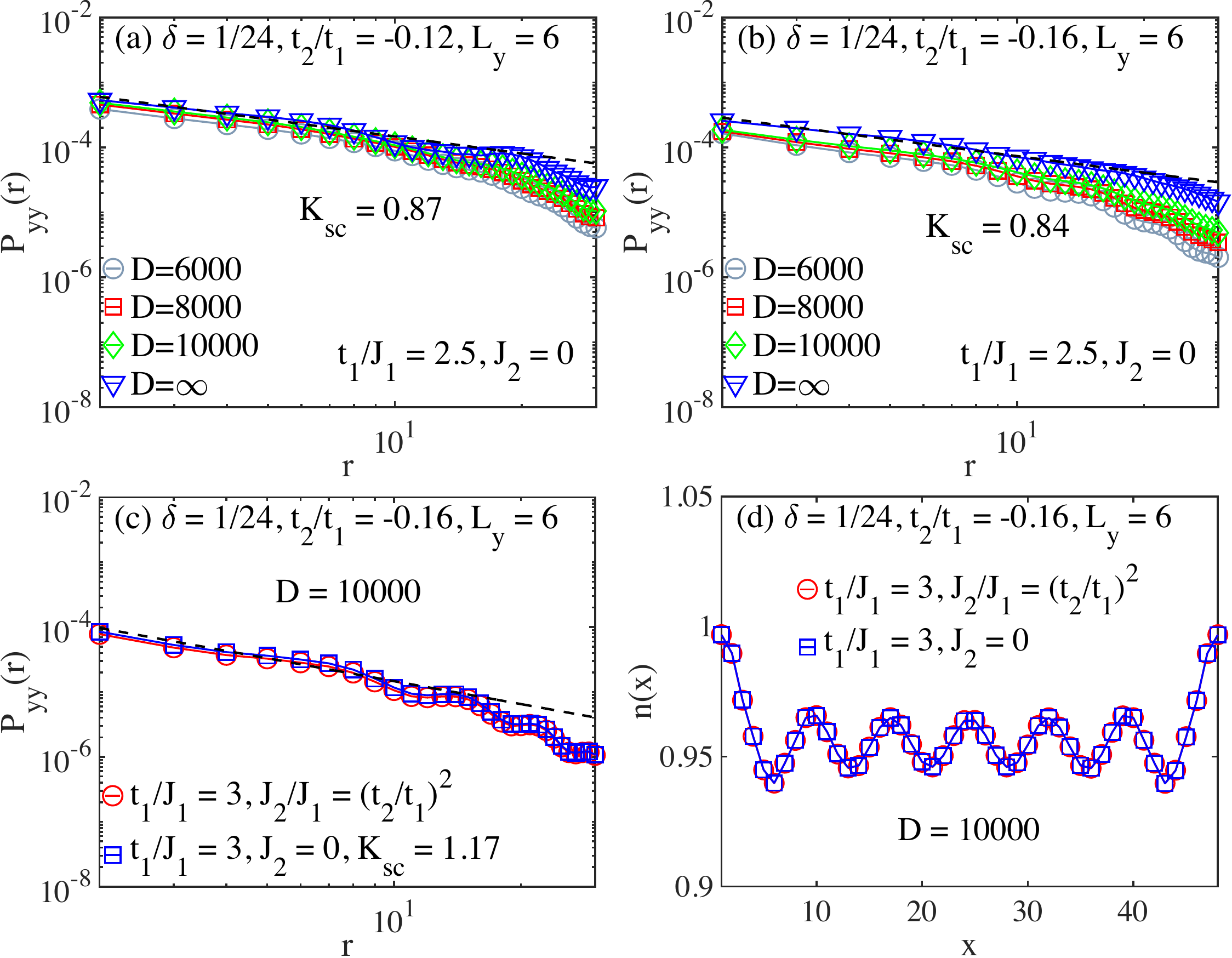}
   \caption{\label{comapre68}
    SC pairing correlation functions $P_{yy}(r)$ for the $t_1$-$t_2$-$J_1$ model with $t_1/J_1 = 2.5$ ($J_2 = 0$) on six-leg systems at $\delta = 1/24$ for (a) $t_2/t_1 = -0.12$ with $K_{\rm sc} \simeq 0.87$, (b) $t_2/t_1 = -0.14$ with $K_{\rm sc} \simeq 0.92$ and (c) $t_2/t_1 = -0.16$ with $K_{\rm sc} \simeq 0.88$.}
\end{figure}

\begin{figure}
   \includegraphics[width=0.7\textwidth,angle=0]{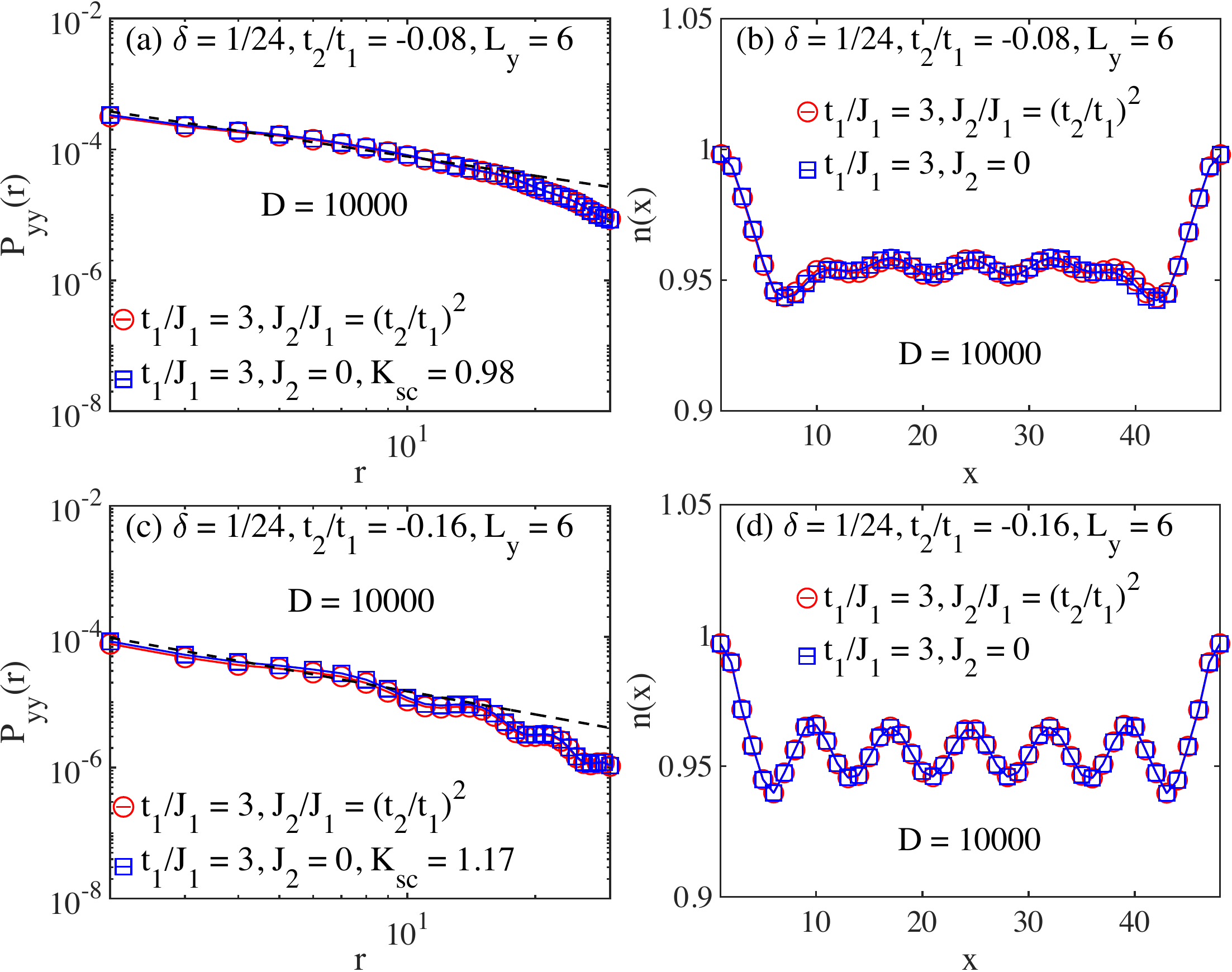}
   \caption{\label{comapreJ2}
     Comparison of the $t$-$J$ model with and without $J_{2}$ interaction on six-leg systems. (a) and (b) compare $P_{yy}(r)$ and $n(x)$, respectively, for $\delta = 1/24, t_2/t_1 = -0.08, t_1/J_1= 3$, $J_2/J_1 = (t_2/t_1 )^2$ and $J_{2}=0$. Here, we directly compare the results obtained with $D=10000$. (c) and (d) are the similar comparisons for $t_2/t_1 = -0.16$.}
\end{figure}

In Fig. 5 of the main text, we have shown the bond dimension extrapolated pairing correlations $P_{yy}(r)$ for $\delta = 1/8$, $t_1/J_1 = 2.5, J_2 = 0$, $t_2/t_1 = -0.1$ [Fig.~5(a)] and $-0.2$ [Fig.~5(c)] on the $L_y = 8, L_x = 24$ cylinder. 
To support the presented data, here we show the bond dimension scaling of the pairing correlations $P_{yy}(r)$ in Fig.~\ref{scaling825}.
The pairing correlations are extrapolated with the bond dimension up to $24000$ SU(2) multiplets.
The pairing correlations grow smoothly with increased bond dimension.
The extrapolated data could be appropriately fitted by a good power-law behavior, giving the power exponents $K_{\rm sc} \simeq 1.37$ for $t_2/t_1= -0.1$ and $K_{\rm sc} \simeq 1.38$ for $t_2/t_1 = -0.2$, respectively.

In Fig. 5 of the main text, we have compared the results of the extended $t$-$J$ models with different parameters, focusing on the effect of the small $J_2$ and different $t_1/J_1 = 2.5$ to the SC phases found on the eight-leg systems.
It turns out that the SC phases are quite robust. 
Here, in Figs.~\ref{comapre68} and \ref{comapreJ2}, we also examine whether the SC + CDW phase found on the six-leg systems would be stable or not at different model parameters. 
One can see that for some different $t_2/t_1$, the pairing correlations all exhibit a good algebraic decay with small power exponents.
Therefore, the SC + CDW phase also can emerge in the $t_1$-$t_2$-$J_1$ model with $t_{1}/J_{1}=2.5$, $J_2 = 0$. 
Besides, we also confirm that the physical properties are almost invariant for $J_2/J_1 = (t_2/t_1)^2$ and $J_2 = 0$, as shown in Fig.~\ref{comapreJ2}.

\section{H. Extrapolation of correlation functions with bond dimension}
\label{sec:bonddimensions}

\begin{figure}
   \includegraphics[width=0.65\textwidth,angle=0]{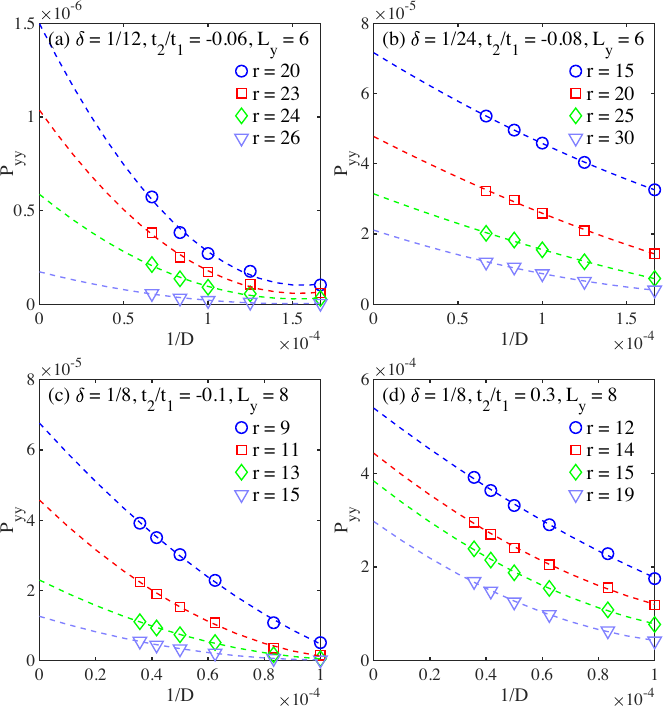}
   \caption{\label{Scaling_details}
   Extrapolations of pairing correlation functions versus bond dimension.
   (a) and (b) show the extrapolations of the pairing correlations $P_{yy}$ on the six-leg cylinder for $t_{2}/t_{1}=-0.06$, $\delta=1/12$ in the stripe phase, and $t_{2}/t_{1}=-0.08$, $\delta=1/24$ in the SC + CDW phase. $D$ is the $\mathrm{SU}{\left(2\right)}$ bond dimension, which corresponds to $D=6000, 8000, 10000, 12000, 15000$. (c) and (d) are the similar extrapolations on the eight-leg cylinder for $t_{2}/t_{1}=-0.1$, $\delta=1/8$ in the hole-doped SC phase, and $t_{2}/t_{1}=0.3$, $\delta=1/8$ in the uniform $d$-wave SC phase. The kept bond dimensions correspond to $D=10000, 12000, 16000, 20000, 24000, 28000$. The correlation data at each given distance $r$ are extrapolated by the quadratic polynomial function 
$\mathcal{F} \left(1/D\right)=\mathcal{F} \left(0\right)+\alpha /D+\beta /D^2$.
}
\end{figure}

In this section, we explain the bond dimension extrapolation for correlation functions in more details.
We perform polynomial extrapolations to best fit the data for a range of bond dimensions up to $15000$ and $28000$ $\rm SU(2)$ multiplets for the six- and eight-leg systems. 
In Fig.~\ref{Scaling_details}, we show the extrapolations of the pairing correlations in the stripe phase (six-leg), SC + CDW phase (six-leg), SC phase (eight-leg), and uniform $d$-wave phase (eight-leg) as representative. 
For each given distance $r$, the correlations are extrapolated by the quadratic polynomial function 
$\mathcal{F} \left(1/D\right)=\mathcal{F} \left(0\right)+\alpha /D+\beta /D^2$ with five or six bond dimension data. 
We have also checked that the cubic polynomial extrapolations give the consistent results.

\end{document}